\documentstyle[12pt,procsla,epsfig]{article}
  
  \def\aprle{\buildrel < \over {_{\sim}}} 
  \def\aprge{\buildrel > \over {_{\sim}}}   
  \def\nubar{\overline{\nu}}

  \catcode`\@=11
  \long\def\@makefntext#1{
  \protect\noindent \hbox to 3.2pt {\hskip-.9pt  
  $^{{\ninerm\@thefnmark}}$\hfil}#1\hfill}		
  \def\@makefnmark{\hbox to 0pt{$^{\@thefnmark}$\hss}}  
  \def\ps@myheadings{\let\@mkboth\@gobbletwo
  \def\@oddhead{\hbox{}
  \rightmark\hfil\ninerm\thepage}   
  \def\@oddfoot{}\def\@evenhead{\ninerm\thepage\hfil
  \leftmark\hbox{}}\def\@evenfoot{}
  \def\sectionmark##1{}\def\subsectionmark##1{}}
  
\begin{document}
  
  \centerline{\normalsize\bf THE PRIMARY PROTONS }
  \baselineskip=16pt
  \centerline{\normalsize\bf AND THE ATMOSPHERIC NEUTRINO FLUXES}

  \vspace*{0.3cm}
  \centerline{\footnotesize PAOLO LIPARI}
  \baselineskip=13pt
  \centerline{\footnotesize\it INFN and Dipartimento di Fisica,
               Universit\`a di Roma ``La Sapienza''}
  \centerline{\footnotesize\it P. A. Moro 2, 00185 Roma, Italy}
  \centerline{\footnotesize E-mail: paolo.lipari@roma1.infn.it}

  \vspace*{0.9cm}

\abstracts{
The predictions  of  the atmospheric $\nu$ event rates 
are affected by significant  uncertainties,
however the  evidence for the `disappearance' of
$\nu_\mu$'s and $\overline{\nu}_\mu$'s 
obtained  by SK  and other underground  detectors
is robust and cannot be  accounted in the framework of the 
minimum  standard  model
without assuming  very large   {\em ad hoc} experimental systematic  effects.
The existence  of `new physics'  beyond the  standard model
is  therefore   close to be established; $\nu$ 
oscillations   provide a  very   good  fit to all data.
The theoretical   uncertainties 
do have  an important  role  in the  detailed interpretation
of the data,   and  in the estimate  of  oscillation parameters.}
   
  \normalsize\baselineskip=15pt
  \setcounter{footnote}{0}
  \renewcommand{\thefootnote}{\alph{footnote}}

\section{Introduction}
The experimental results of  Super--Kamiokande (SK)\cite {SK,SK-new} and other 
detectors\cite{Kamioka,IMB,Soudan,MACRO}  on    
atmospheric  neutrinos have  provided  strong evidence 
for the  existence of $\nu$   oscillations   or other forms
of   `new phyics'  beyond  the minimum standard model.
This  conclusion  comes  from  the comparison of 
the  experimental  results with theoretical predictions  
that have   schematically    the  structure
(for  $\mu$--like  events as  an example):
\begin {equation}
{d N_\mu^{\rm th} \over dE_\mu \; d \Omega_\mu} 
 =  \sum_{\alpha=e,\mu} \; 
[ \phi_0 \otimes Y_{p,n \to \nu_\alpha} 
\otimes  P(\nu_\alpha \to \nu_\mu) ] \otimes \sigma_\nu
\otimes A_{det}
\end {equation}
where $N_\mu^{\rm th}$ is  the  predicted  $\mu$ event rate,
$E_\mu$ and $\Omega_\mu$ are the muon energy and  direction, 
and the different   ingredients of the prediction  are:
(i) $\phi_0$ the primary cosmic ray (c.r.) flux;
(ii) $Y_{p(n) \to \nu_\alpha}$ the  $\nu$  yield, that is 
the  number of neutrinos
of flavor $\alpha$  produced  in  $p$ ($n$)  induced  c.r. showers;
(iii)  $\sigma_\nu$ the  neutrino cross sections;
(iv) $P(\nu_\alpha \to \nu_\beta)$  the oscillation   probability;
and (v) $A_{det}$  the  detector acceptance.
We have dropped  all   dependences on flavor, energy  and  zenith angle
of the different elements  and the symbol  $\otimes$ indicates the need
for   appropriate  convolutions   that   can   be done  accurately only
with  Montecarlo methods.

The  fundamental result  
obtained  with the   measurements of atmospheric  $\nu$'s 
is that the data {\em cannot}  be described in the minimum  standard model, 
(in the absence of oscillations).
In particular    the $\nu_\mu$ 
survival  probability $P(\nu_\mu \to \nu_\mu)$ must be  smaller
than unity  in   a  large region of energy and $\nu$ pathlenghth.
In short: muon neutrinos  disappear.

It is  possible  to go beyond  this  fundamental but qualitative
statement.  With the inclusion of $\nu$--oscillations
one can  obtain  a good description of  the data and   
extract  quantitative 
information on the  $\nu$ masses and  mixing\footnote{
Other mechanisms, different from  oscillations, such  as $\nu$ decay,
FCNC,  or violations  of the equivalence  principle,  have been proposed  
as explanations  of the disappearance of  the  muon neutrinos.
These alternative or `exotic'  models  provide  (at least until now)
significantly  less satisfactory description of the data.
See\protect\cite{Pakvasa,exotic} for  more discussion.}.
Two  flavors $\nu_\mu \leftrightarrow \nu_\tau$ oscillations  
with  maximal  of very  large mixing  are
a  viable explanation  for  the data.
The determination of the $\nu$ oscillation  parameters
is of course of fundamental importance, and has also a crucial role to
determine the   best  strategy for  future experimental  studies.

In this  work we will analyse  the first three elements of the
predictions: the primary  flux,  the modeling of hadronic  showers  and
the $\nu$ cross sections
(without  considering at all  the important questions  of 
detector  acceptance and background estimates),
and discuss what are the  effects of
the existing uncertainties.
We will  argue that  the   fundamental 
result  that the minimum  standard  model  fails to describe the
atmospheric $\nu$  data is   robust, and  that  the 
evidence for  the disappearance of $\nu_\mu$'s  cannot be 
``reabsorbed''  taking  into account theoretical  uncertainties.
On the  other hand these  uncertainties 
can play a  significant role
in the interpretation of the data  and in the  determination
of the oscillation  parameters.

In this  work  we  will    refer to  some published   calculations
of the atmospheric $\nu$ fluxes.
Two detailed  Monte Carlo (MC) calculations of  the  
fluxes performed by the
Bartol  group\cite{Bartol} and  by 
Honda, Kajita, Kasahara and Midorikawa (HKKM)\cite{HKKM}
have  been used to  compute    predictions 
for  the  interpretation of the existing  data.
Other  calculations  
have  been   performed  analytically by  Bugaev and Naumov\cite{BN}
or  again with MC methods by Lee  and Koh\cite{LK}.
A new  detailed  calculation  using the FLUKA
code\cite{fluka} for the modeling of hadronic interaction
is  close to completion\cite{nu-fluka}.

This   discussion is organized  as  follows.
In the next section we  will briefly  discuss
the evidence  for new physics in the
atmospheric  $\nu$ data; 
in section 3  we will describe  some   predictions
for  the atmospheric $\nu$ fluxes;
in section 4, 5 and 6
we will consider the three main elements in the prediction:
the  primary cosmic  ray  flux,
the development of  hadronic showers;
and the $\nu$ cross sections;
in section 7 we  will consider the  constraint  coming  from
atmospheric  muon measurements;
in section  8 we discuss   discuss the question of the absolute
normalization  of the  $\nu$ flux; in section  9
we  will   discuss  some possible  effects 
of the uncertainties in the predictions  for the
determination of the oscillation parameters.

\section {Evidence for  the  disappearance of $\nu_\mu$'s}
The   evidence  for the disappearance of muon 
(anti)--neutrinos comes  from the  observation of  three experimental effects,
listed here in order of `robustness':
(i)  the detection  of an up--down asymmetry
for the $\mu$--like  events,
(ii) the detection of  a small $\mu/e$ ratio,
(iii) the detection of  a  distortion   of the zenith angle
distribution and  a suppression of the $\nu$ induced upward 
going muon  flux.

Some  comments  about the first two  effects  can be  made already before  
entering in a detailed  discussion of the prediction
because two  properties  of  the $\nu$ fluxes:
an approximate  up--down  symmetry, and   a strongly
constrained $\nu_\mu/\nu_e$ ratio,
are to a large extent  independent from the details  of the 
calculation   and  provide  `self calibration'    methods.

\vspace {0.25 cm}
\noindent (i) The $\nu$ fluxes are predicted  
in the no--oscillation  hypothesis  to  be 
approximately up/down  symmetric:
\begin{equation}
\phi_{\nu_\alpha}  (E_\nu, \theta)   \simeq  
\phi_{\nu_\alpha}  (E_\nu, \pi - \theta)
\end{equation}
This  follows  as  a simple 
and {\em  purely geometrical}  theorem    from two  assumptions:
(a) the  primary c.r.  flux  is isotropic, 
(b) the Earth is  spherically symmetric.
The geomagnetic   field, as will be discussed in some  detail
in section 4.2,   is the  most important source of up--down 
asymmetry.
The theory predicts unambiguosly that the effects  of the   geomagnetic field
are small,  decreasing with   $E_\nu$  and for  Kamioka
produce an excess of {\em  up--going} particles.
This  predictions  have  received an important
quantitative  experimental confirmation  with the measurement of 
the east--west effect in SK\cite{SK-east-west},  that is also
produced by the geomagnetic field, and is  largely  independent
from oscillations\cite{geom}.
For the  sub--GeV and multi--GeV  muon  samples   in SK
the up/down  asymmetry $A = { (U-D) / (U + D)} $
(where $U$  ($D$) is the number of up (down)  going events with
$\cos\theta_\mu \le -0.2$ ($\cos\theta_\mu \ge -0.2$)
takes the  values:
\begin{equation}
A_{sub} = -0.150 \pm 0.028 \pm 0.01   ~~~~~
A_{mul} = -0.311 \pm 0.043 \pm 0.01
\end{equation}
these  are deviations  of 5.4  and 7.2 standard deviations  from
no asymmetry.

\begin{figure} [bt]
\centerline{\psfig{figure=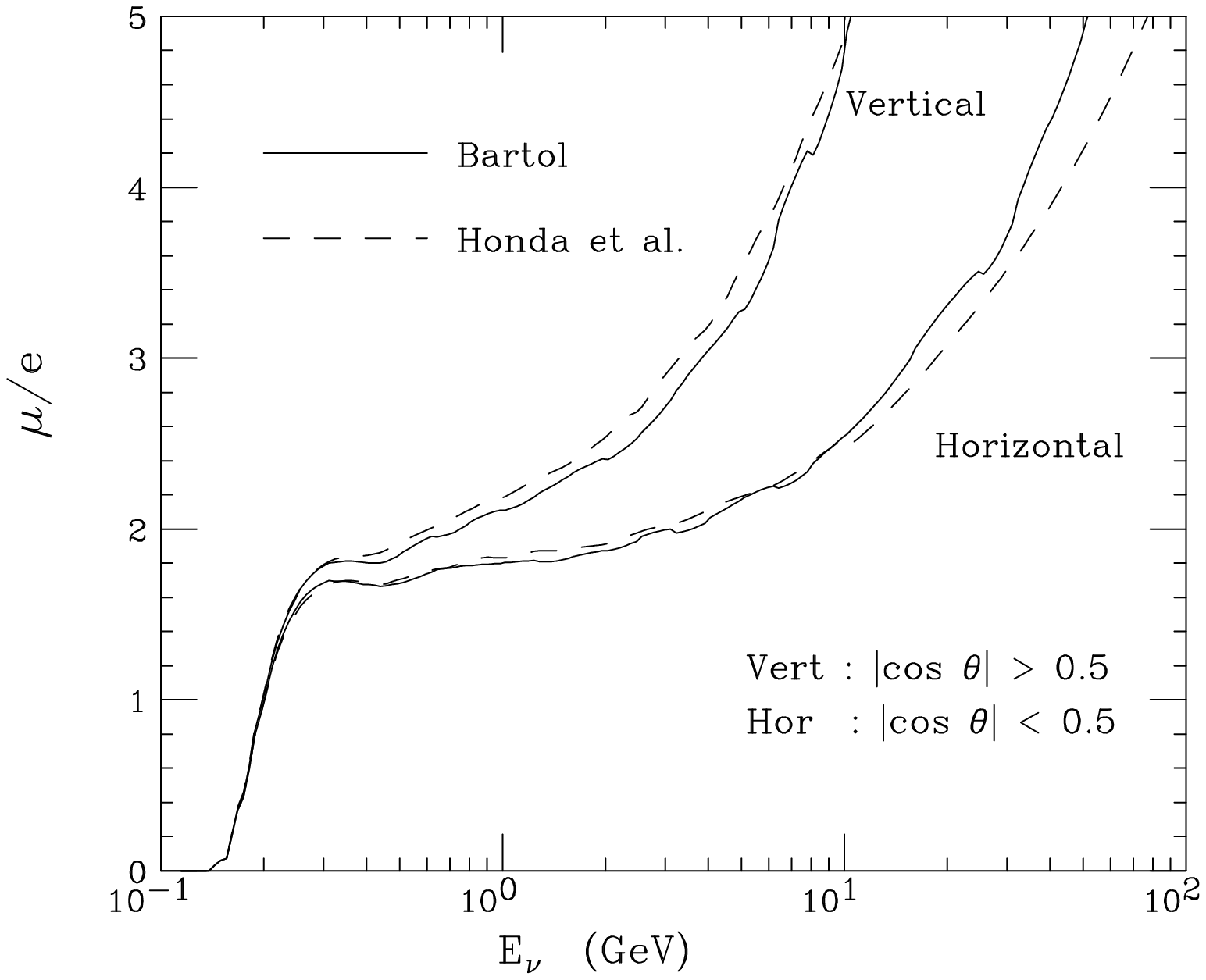,height=9cm}}

\vspace{0.3 cm}
\fcaption {$\mu/e$ ratio     of
cc--interacting $\nu$'s  calculated in the
no--oscillation hypothesis  using the
 Bartol \protect\cite{Bartol}  and 
 HKKM \protect\cite{HKKM}   fluxes.
\label{fig:r}}
\end{figure}

\vspace {0.25 cm}
\noindent (ii) The  $\nu_\mu$  and $\nu_e$ fluxes
are strictly related  to each other  because they are
produced  in the   chain decay of the same
charged  mesons (as in $\pi^+ \to \nu_\mu \, \mu^+
\to \nu_\mu \,(\overline{\nu}_\mu \nu_e e^+$)):
\begin{equation}
\phi_{\nu_\mu} (E,\theta) = r_\nu (E_\nu, \theta) \times \phi_{\nu_e} (E,\theta)
\end{equation}
the  factor $r_{\nu(\nubar)} (E_\nu, \theta)$  vary  slowly with
energy and  angle  is quite insensitive   to   the details  of 
the calculation. 
At low energy   when  essentially all  parent  muons
decay before reaching the   ground one has  
$r_{\nu(\overline{\nu})} \simeq 2$.
This is  in  agreement with a 
naive $\nu$ `counting'  argument.
This  simple   argument   is    approximately valid
because   after chain decay
the three neutrinos   produced  in the decay of
a pion have  similar average energy
(for ultrarelativistic 
$\pi^+ \to \nu_\mu(\nubar_\mu \nu_e e^+)$ decay one has:
$\langle E_{\nu}/E_\pi \rangle = 0.213$,
0.265  and 0.257  for the $\nu_\mu$ $\nubar_\mu$ and $\nu_e$ channels).
With increasing energy, because of the Lorentz  time  dilatation,
the muons begin  to reach the ground  before  decay  and 
$r_{\nu(\overline{\nu})}$ increases, with  a  stronger  (weaker) effect
for  vertical  (horizontal)   directions  because  of the differen
distance between the primary interaction point  and the ground.
These energy and zenith angle  dependence  of $r$ (see fig.~\ref{fig:r})
are  controlled  essentially by  geometry  and   reliably calculable.
In SK the the double  ratio
$R = (\mu/e)_{data}/ (\mu/e)_{MC} $
 for the sub--GeV and multi--GeV  samples are  measured as:
\begin{equation}
R_{sub} = 0.668 {~}^{+0.026}_{-0.023} \pm 0.007 \pm 0.052  ~~~
R_{mul} = 0.663 {~} ^{+0.044}_{-0.041} \pm 0.013 \pm 0.078
\end{equation}
where the errors are  statistical, systematic   and theoretical
(as  estimated by SK).
Combining in quadrature all errors  the significance  of the deviations
of the  double  ratios  from unity are 5.7  and  3.7  standard  deviations.
The  uncertainty in the denominator  of the double ratios  has  several
sources: the ratio  $\pi^+/\pi^-$  (that  determines  
ratio $\nu_e/\nubar_e$  important because of the different
cross sections  of  $\nu$'s  and $\nubar$'s),
the contribution of  kaons\footnote{In the chain decay
$K^+ \to \nu_\mu \mu^+ \to \nu_\mu (\nubar_\mu \nu_e e^-)$
the average  energy of the three neutrinos is 
$\langle E_\nu/E_K \rangle = 0.477$, 0.159  and 0.205,
therefore for   neutrinos from $K$ decay the naive 
$\nu$ `counting'  argument is  not  valid.
This  is  compensated by the fact that 
charged ad  neutral kaons are also  a  source of  electron neutrinos via 
the decays $K \to \pi e\nu_e$.}, the   shape of
the primary spectrum,    and the  momentum distribution  of the produced
mesons. 
These  uncertainties  have estimated\cite{SK,atm-comp,Gaisser} 
as  of order 5\%  for  a   fixed   value of $E_\nu$.
Since experimentally the $(\mu/e)$ implies   an integration over
a  finite  interval of  energy
and the  convolution of the  $\nu$ specra with 
detection   efficiency curves  that are  flavor  dependent,
an important  source of  uncertainty comes  from  the  knowledge of the
shape of the energy spectrum, and the energy  dependence of the
$\nu$  cross section. For this  reason   the uncertainty on
$(\mu/e)_{MC}$  is  larger than  5\%
and  was  estimated by SK as
8\% and 12\% for the sub--GeV and multi--GeV sample.
The   estimate   of the value  and uncertainty
of the denominator
in  the  double ratios  is  very important  for  the interpretation
of the data. In any case 
even  with  a  very  conservative  
estimate  of `theoretical errors'
it does  not  appear possible to
explain   the detected low  values of $R$ obtained  without
appealing to {\em experimental} errors  in flavor  identification.

\vspace {0.25 cm}
\noindent (iii)  For the interpretation of the upward--going  muon 
data,  the  methods  of comparing  up and down hemispheres,
or the $\nu_\mu$ and $\nu_e$   fluxes are  not available, 
therefore the    uncertainties in the  predictions are 
more important.
The  angular distribution of the  muons,
and the ratio   between a  high  energy and a low energy 
part  of the  signal  (or example stopping and  through--going muons),
are  less sensitive  to  the model  (see \cite{ll-upmu} for a  
critical discussion), and provide 
important tools  for  the interpretation of the data\cite{upmu}.
Upward going muons are the best way to  study the $\nu_\mu$  fluxes
at high  $E_\nu$,  and this is important  to study experimentally 
the energy dependence  of the  $\nu$ flavor   transitions\cite{exotic}.

\section {Neutrino  Fluxes  calculations}
In this section  we will   briefly  present some  results 
 about the  properties of the calculated  neutrino  fluxes.
For  the sake of  clarity,   we  will  discus
the   $\nu$   event rate   
that (for example for  $e$--like events) can be calculated as:
\begin {equation}
{d^2 N_{e} \over dE_\nu \,d \cos\theta_\nu} =
 N_A ~
~[\phi_{\nu_e} (E_\nu, \cos \theta_\nu) \;\sigma_{\nu_e}^{CC} (E_\nu) + 
  \phi_{\nubar_e} (E_\nu, \cos \theta_\nu) \;\sigma_{\nubar_e}^{CC} (E_\nu)]
\end{equation}
as  the product of the  flux times the  neutrino
cross sections.  For $\sigma_\nu$ we  will use
in this   work  the  estimates of \cite{LLS}  including
nuclear  effects.

\begin{figure} [bt]
\centerline{\psfig{figure=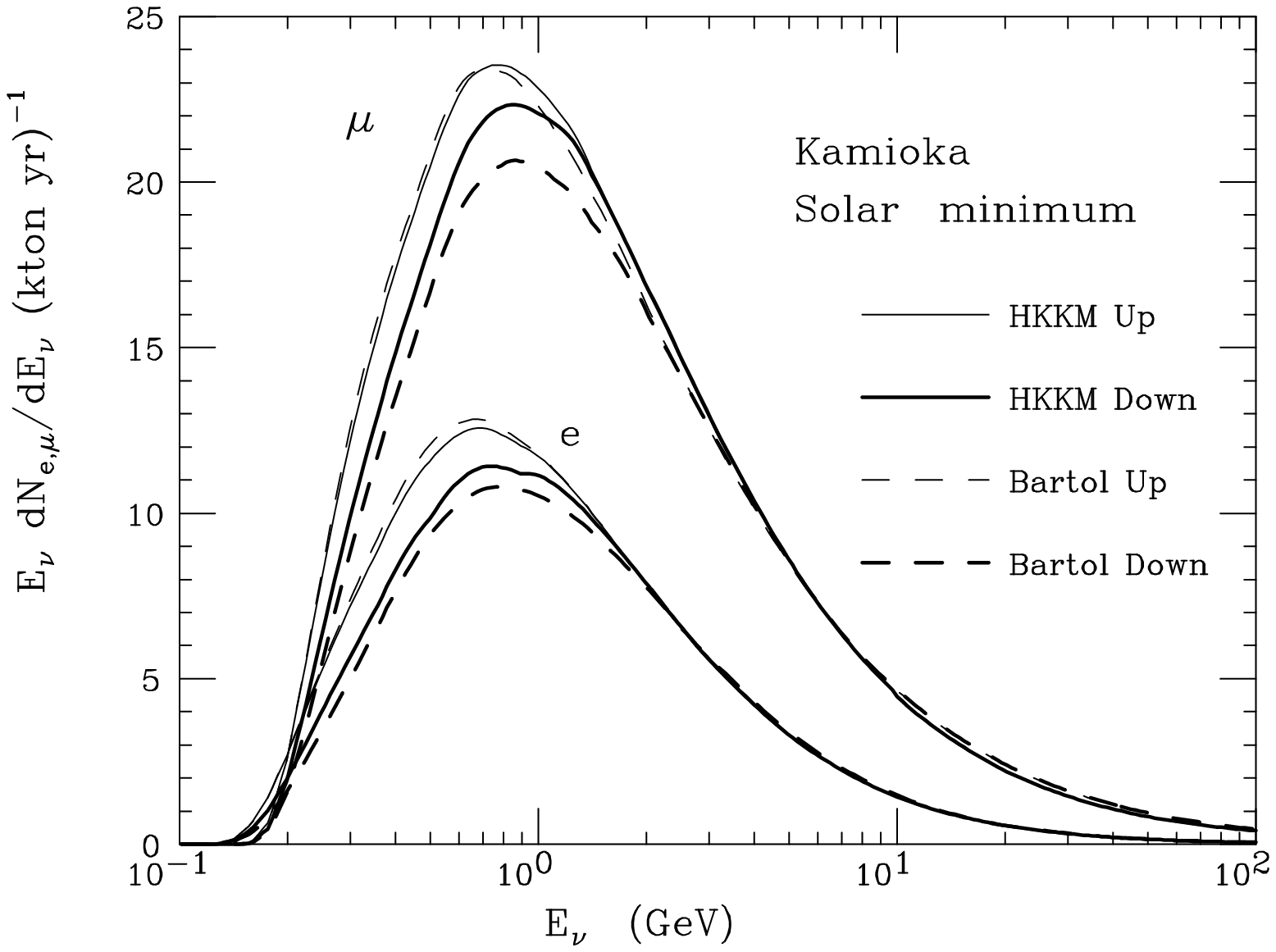,height=9cm}}
\fcaption {Energy  distributions
of   up--going  ($\cos\theta_\nu <0$)
and   down--going  ($\cos\theta_\nu >0$)  
cc--interacting 
$(\nu_e +\nubar_e)$'s  and 
$(\nu_\mu +\nubar_\mu)$'s   at Kamioka
The calculations  are    for solar  minimum 
and  with the Bartol\cite{Bartol} and HKKM\cite{HKKM}
 $\nu$ fluxes.
\label{fig:kam_rate}}
\end{figure}

In fig.~\ref{fig:kam_rate} 
we  show  the energy distributions of   up--going
and  down--going  interacting  neutrinos  at Kamioka 
as calculated    for   minimum solar activity,
by  the Bartol\cite{Bartol} an HKKM\cite{HKKM}   groups.
The $\nu$  direction  was integrated  
over    the angular  regions:
$\cos \theta_\nu \in [-1,0]$ for  `Up' events,
or $\cos \theta_\nu \in [0,1]$ for  `Down' events.
Plotting $E_\nu dN/dE_\nu  = dN/d\ln E_\nu$   with  a
logarithmic  energy scale, the  area  below the curve is  proportional
to the total   rate.
Some  remarks  about fig.~\ref{fig:kam_rate}:

\noindent (i) The two  calculations, especially
for $E_\nu \aprge 1$~GeV, are   quite  similar to each
other  both  in   absolute normalization and   in  the 
shape of  energy distribution.  
This similarity, as it will be discussed  later,
is  to some extent the  result of a
cancellation  of  a higher   (lower) primary cosmic  ray flux
and  a lower (higher)  $\nu$  yield  per primary particle
for the HKKM (Bartol)  calculation.

\noindent (ii) 
Both calculations   predict that  the up--going  and down--going
$\nu$ fluxes   become  equal   for  $E_\nu \aprge 1$~GeV.
At low  energy 
(and in the absence of oscillations) 
the  calculations   predict   a    higher    up--going flux.
This  is the result of the geomagnetic  effects
 and has {\em opposite} sign  with  respect to the effect
of oscillations. 
It should be noticed  that the Bartol  calculation predicts a
stronger   no--oscillation up--down asymmetry  for neutrinos
with $E_\nu \aprle 1$~GeV.

\vspace {0.25 cm}
The calculated  $\nu$  event rates   depends on
the detector position and the epoch of  solar  activity.
This is  illustrated 
in table~1 and~2,  that   give  the 
charged  current  event rates 
(with  no  detector    efficiency
or requirement   for containement included)
for  two  detector  positions 
(the Kamioka  and Soudan mines)  and different levels of
solar  activity.  For the   Kamioka site, 
we also compare the two 
MC calculations  of Bartol and HKKM.
The    quantities    ($U_L$, $U_H$, $D_L$ and  $D_H$)
refer  to  an integration  in the $\nu$  zenith  angle in the regions:
$\cos \theta_\nu \in [-1,0]$ for  `Up' events,
or $\cos \theta_\nu \in [0,1]$ for  `Down' events,
and 
an integration  in   energy 
$E_\nu \le 1$~GeV  for `Low'  energy and
$E_\nu \ge 1$~GeV  for `High'  energy.
The asymmetry is  calculated as $A = (U-D)/(U+D)$.

\begin{table}[hbt]
\tcaption{Event rates  (kton yr)$^{-1}$
 and Up/Down asymmetry for $e$ like event.}
\begin{center}
\begin {tabular} {| l l c |  c c c | c c c | }
\hline
 Site & Flux &  $\odot$ & $U_L^e$ & $D_L^e$ & $A_L^e$ & 
     $U_H^e$ & $D_H^e$ & $A_H^e$ \\
 \hline
 Kamioka  & HKKM & min  &
    15.8 &   13.5 &  0.077 &   14.5 &   14.4 &  0.002 \\
 Kamioka & HKKM & mid  &
    15.1 &   13.1 &  0.070 &   14.2 &   14.2 &  0.001 \\ 
 Kamioka & HKKM & max  &  
    14.3 &   12.7 &  0.060 &   13.9 &   13.9 &  0.000 \\ 
\hline
 Kamioka & Bartol & min  &
    16.2 &   12.5 &  0.129 &   14.6 &   14.2 &  0.013 \\ 
 Kamioka & Bartol & max  &
    14.3 &   12.2 &  0.077 &   14.3 &   14.1 &  0.006 \\ 
\hline
 Soudan & Bartol & min  &
    20.7 &   28.4 & $-0.158$ &   16.3 &   17.6 & $-0.039$ \\ 
 Soudan & Bartol & max  &
    17.5 &   21.9 & $-0.112$ &   15.8 &   16.8 & $-0.031$ \\ 
 \hline
\end{tabular}
\end{center}
\end{table}

\begin{table}[hbt]
\tcaption{Event rates  (kton yr)$^{-1}$
 and Up/Down asymmetry for $\mu$ like event.}
\begin{center}
\begin {tabular} {| l l c |  c c c | c c c | }
\hline
 Site & Flux &  $\odot$ & $U_L^\mu$ & $D_L^\mu$ & $A_L^\mu$ & 
     $U_H^\mu$ & $D_H^\mu$ & $A_H^\mu$ \\
 \hline
 Kamioka & HKKM& max  &
    27.5 &   24.4 &  0.059 &   34.3 &   34.2 &  0.001 \\
 Kamioka & HKKM & mid  &
    26.5 &   23.8 &  0.053 &   33.8 &   33.8 &  0.000 \\ 
 Kamioka & HKKM & max  &  
    25.2 &   23.0 &  0.046 &   33.1 &   33.2 &  0.000 \\ 
\hline
 Kamioka & Bartol & min  &
    27.8 &   22.3 &  0.109 &   33.9 &   33.4 &  0.008 \\ 
 Kamioka & Bartol  & min  &
    25.1 &   21.9 &  0.067 &   33.5 &   33.3 &  0.004 \\ 
\hline
 Soudan & Bartol & min  &
    34.1 &   44.8 & $-0.136$ &   36.4 &   38.4 & $-0.026$ \\ 
 Soudan & Bartol & max  &
    29.8 &   36.5 & $-0.100$ &   35.8 &   37.3 & $-0.021$ \\ 
 \hline
\end{tabular}
\end{center}
\end{table}

\noindent Some remarks can be useful:

\noindent (i) For  sufficiently large  $E_\nu$
 ($E_\nu \aprge  1$~GeV)
 the calculated  event rate   is: (a) independent from  energy,
 (b)  independent from the detector  position, and (c)
 up/down  symmetric.  Conversely  at low energy 
 there  is  a significant  dependence  on detector location
and solar epoch,  and there are  significant  deviations
 from  up/down   symmetry  even in the absence of  oscillations.

\noindent (ii) At the epoch  of solar minimum, 
that to a  good approximation   applies 
to the data  taking  period of SK  (see fig.~\ref{fig:n_monitor}),
without considering the effect of detector efficiency,
the rates of $e$--like  events
with  $E_\nu \le 1$~GeV ($E_\nu \ge 1$~GeV)
 in Soudan and SK 
are   predicted by the Bartol calculation 
to be in the  ratio:  Soudan/Kamioka $\simeq 1.72$ (1.17).
The experimental  results suggests a lower ratio
(see section~7 for more  discussion).

\noindent (iii) All  event rates  should  be  reduced during the
future  period of solar  maximum, with   a  significantly
larger reduction for  Soudan  because of the lower  geomagnetic cutoff.
As an illustration, the  ratio
($\odot$~max)/($\odot$~min) for sub--GeV $e$--like  events
is  predicted    by the Bartol calculation to be
$\sim 0.80$ for  Soudan  and $\sim 0.93$  for  Kamioka.
These  time  variations  should be measurable and offer
a handle for a check of the predictions.

\begin{figure} [bt]
\centerline{\psfig{figure=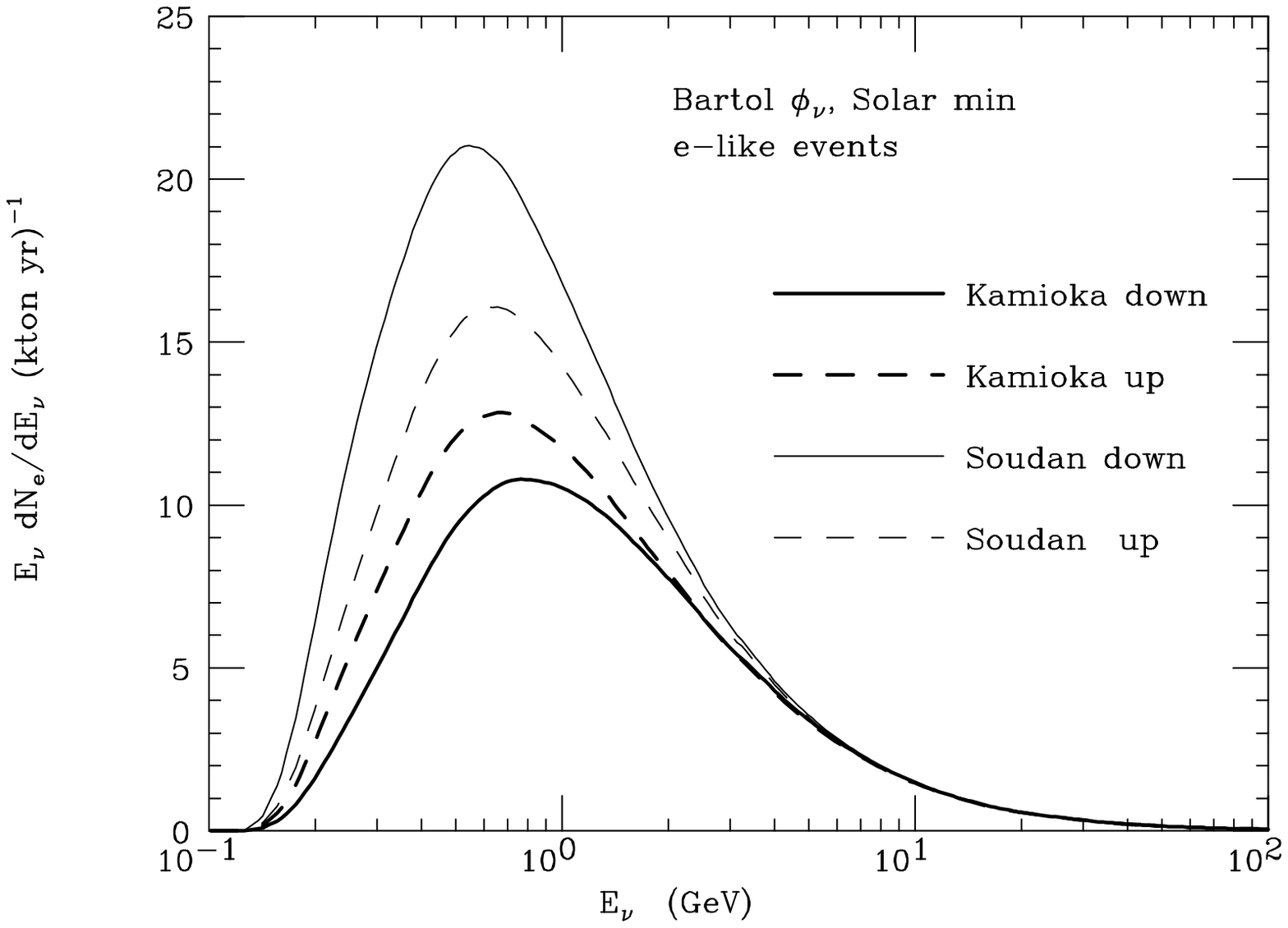,height=9cm}}
\fcaption {Energy  distributions
of   up--going  ($\cos\theta_\nu <0$)
and   down--going  ($\cos\theta_\nu >0$)  
interacting $\nu_e$'s and $\nubar_e$'s   at Kamioka
and Soudan. 
The calculations  are  based  on the solar  minimum 
Bartol $\nu$ fluxes.
\label{fig:nu_rate}}
\end{figure}

\noindent (iv) At low  energy the geomagnetic effect   result in a significant
up--down asymmetry even in the absence  of oscillations.
It is remarkable that this asymmetry 
has  opposite signs  at Soudan  (with an  excess of 
down--going  $\nu$'s)  and Kamioka
(excess of up--going $\nu$'s).  This  effect is also
illustrated in fig.~\ref{fig:nu_rate}.
\noindent The difference  between  Soudan  and Kamioka 
can be easily    understood qualitatively   taking into account the
fact that the  magnetic latitudes of  Kamioka  and Soudan
are  $\lambda_M \simeq 27^\circ$  and $56^\circ$, and the
different  geomagnetic  cutoffs for  the two  locations
(see section 4.2).

\subsection {The response  function}
\begin{figure} [bt]
\centerline{\psfig{figure=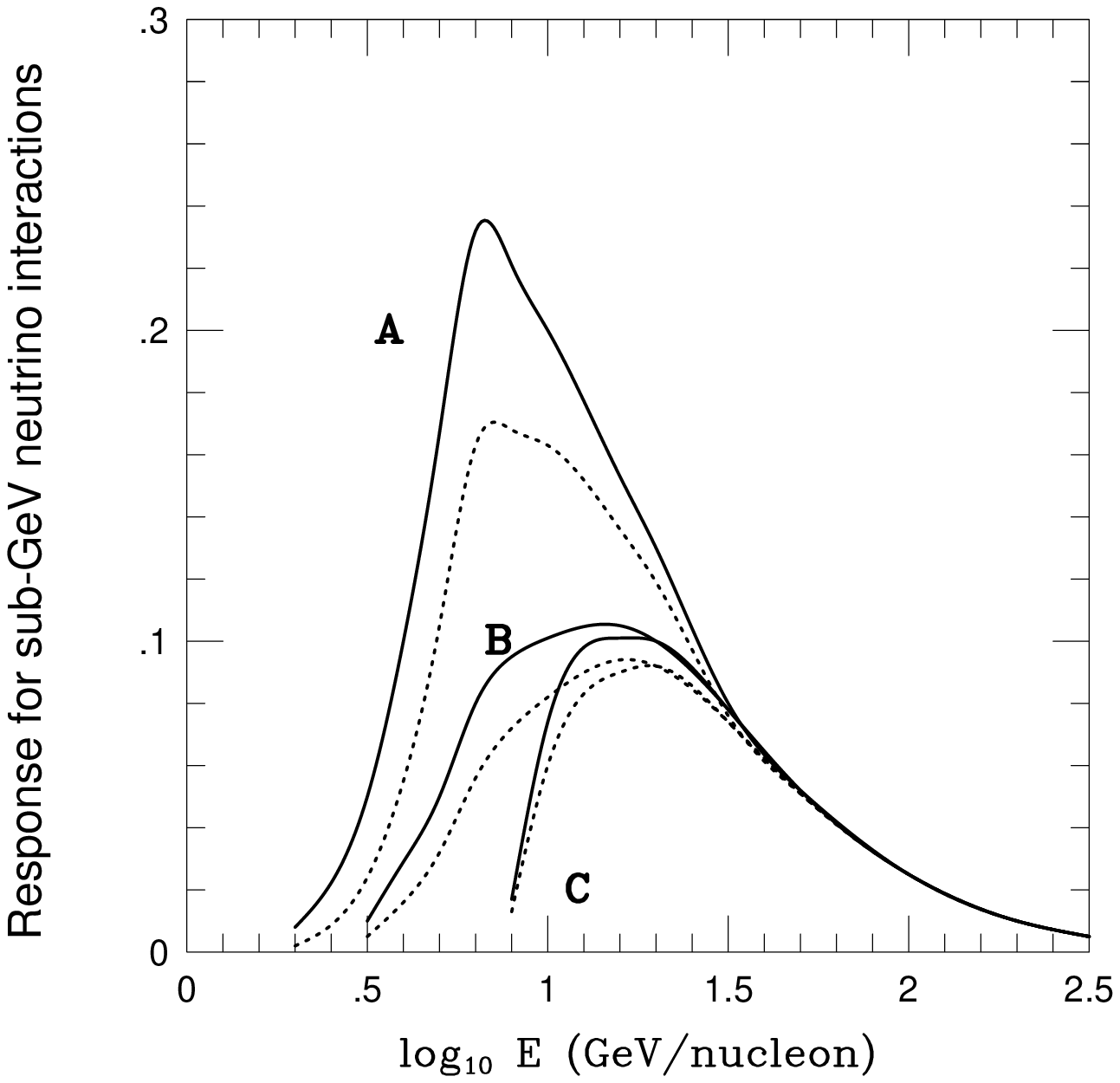,height=9cm}}
\fcaption {From   Gaisser\cite{Gaisser}.
Energy distribution of primary particles  that
produce the   sub--GeV $\mu$--like event rate in
Super--Kamiokande  (see text). 
\label{fig:response}}
\end{figure}

Fig.~\ref{fig:response}   is  taken from 
Gaisser\cite{Gaisser} and shows the  energy distribution
of  the primary particles that (would) contribute to  the 
to the $\mu$--like sub-GeV signal in Super-Kamiokande.
The solid (dotted) curves are 
for solar minimum (maximum).
The pair  of  curves  indicated  as B  (C)  are the 
response  curves for up--going (down--going)   neutrinos  at 
Kamioka.    One can notice    how the lower
flux  of the down--going  emisphere
originates from the   higher  geomagnetic  cutoffs
relevant   for the low  magnetic  latitude site of Kamioka.
Curve A is calculated  assuming zero  cutoff  and
is  a  reasonable  approximation for the 
down--going   $\nu$ flux   of the high magnetic  latitude
of  Soudan  where  geomagnetic cutoffs are small.
The  primary energies that are important  
for  the   atmospheric  $\nu$ signal   are  in a  broad range
$E_0 \sim 3$--100~GeV.  It  is  also important to  notice
that experiments   made in different  locations
will be  sensitive to  somewhat different   ranges 
of  primary energy, and that in particular  the  
event rates at Soudan receive a  large  contribution,
from particles with $E_0 \simeq 10$~GeV, and 
the   prediction   is  very sensitive to the 
flux and  interaction  properties of   low  energy primaries.

\section {The  primary cosmic ray flux}
\begin{figure} [bt]
\centerline{\psfig{figure=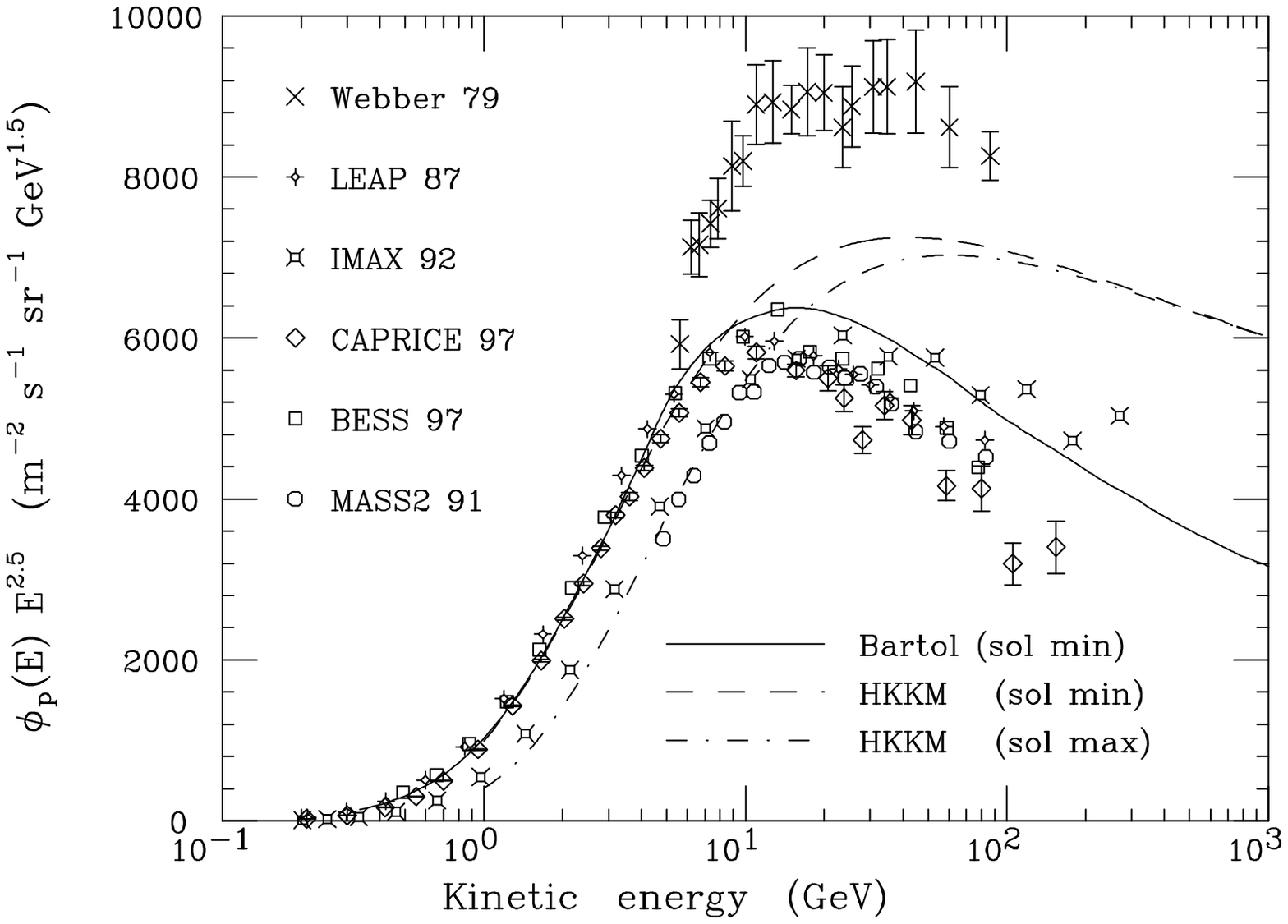,height=9cm}}
\fcaption {Measurements of the c.r. proton flux  (points)
and the representations
used as  input in the  Bartol 
and HKKM  calculations (lines).
\label{fig:proton}}
\end{figure}

The  primary cosmic  ray  flux  has  been  a major  source
of  uncertainty  (see\cite{Gaisser} for a detailed  discussion)
because of   the discrepant  results
 obtained  by two  groups: Webber\cite{Webber} and LEAP\cite{LEAP}),
differing by  $\sim 50\%$ (see  fig.~\ref{fig:proton}).
In  more recent time  there  have    been four
new  measurements  of the $p$ flux:
IMAX\cite{IMAX}, CAPRICE\cite{CAPRICE},
BESS\cite{BESS} and MASS2\cite{MASS2}) of the    proton   flux
(some  including also  measurement of higher mass primaries).
All four   these new  measuremens (shown in fig.~\ref{fig:proton}) are, 
at least qualitatively,  in agreement between  each other and
confirm  the lower  normalization of the LEAP\cite{LEAP}   experiment.
The conclusion  is  that, if one 
 is  ready to   discard  the early measurement
of  Webber\cite{Webber},  the primary flux  is  much  better
determined,  with an uncertainty  in the  relevant region of
better than 10\%.

\begin{figure} [bt]
\centerline{\psfig{figure=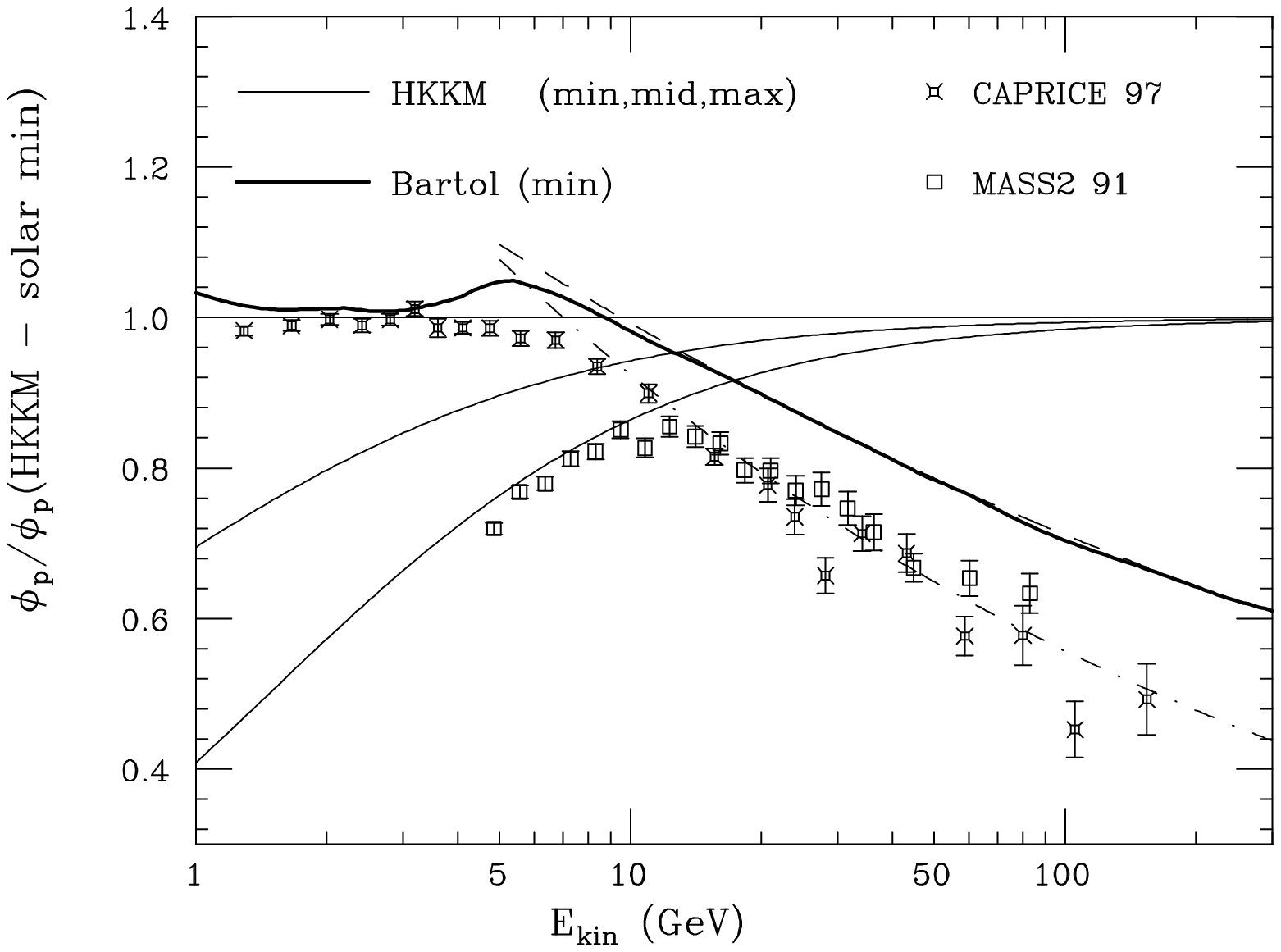,height=9cm}}
\fcaption {Comparison of   measurements of the primary
c.r.  flux  and   representations  used in calculation
of the $\nu$--fluxes (see text).
\label{fig:comp}}
\end{figure}

It is  important to   compare the new experimental  results
on the $p$ flux with the 
representations
chosen for the calculations of the  $\nu$ fluxes.
This  comparison is shown in fig.~\ref{fig:proton} and in more  detail  in 
fig.~\ref{fig:comp}.  
where the data  and   the   different  descriptions
of the $p$ flux   are   shown    as a ratio
with respect   to  the  proton flux   used   in the
HKKM\cite{HKKM}   calculation for minimum  solar  activity.
The  three thin solid  lines   describe  the 
$p$ flux    assumed  for    by HKKM  for 
minimum, medium and maximum solar activity 
(upper  line  with constant value  of unity, middle line and  lower line).
The  thick  line is  the description of  the $p$ flux  
at solar minimum  used in the Bartol  calculation.
Only two sets of data points  are shown,
those obtained  by the Caprice\cite{CAPRICE}  and MASS2\cite{MASS2}
experiments. These  data  were  obtained  in 1997 and 1991
approximately  at the epoch  of minimum  (maximum)   solar  activity.
Some  remarks:

\noindent (i) The   measurements  are consistent   with each other  
for $E_p \aprge 10~GeV$.  Below this  energy  the difference
in   the measured flux, agrees with  the expectations
for the  solar modulation.

\noindent (ii) For $E_p \aprge 10$~GeV,  the descriptions of the  $p$  flux
used in  the calculations  of the $\nu$ fluxes
are in poor  agreement with  the data.
The  discrepancy  is   {\em  not}    a constant factor.
As  an illustration  in fig.~\ref{fig:comp}
  we have  included  a dot--dashed
line: $(E_p$/7~GeV)$^{-0.22}$, that 
corresponds  very  roughly to the ratio  data/model  for  
the solar minimum  $p$  flux of HKKM,
and a dashed  line
($E_k$/9.5~GeV)$^{-0.144}$), that    corresponds roughly to the 
ratio Bartol/HKKM  at  solar  minimum.
 
\subsection {Time  dependence}
\begin{figure} [bt]
\centerline{\psfig{figure=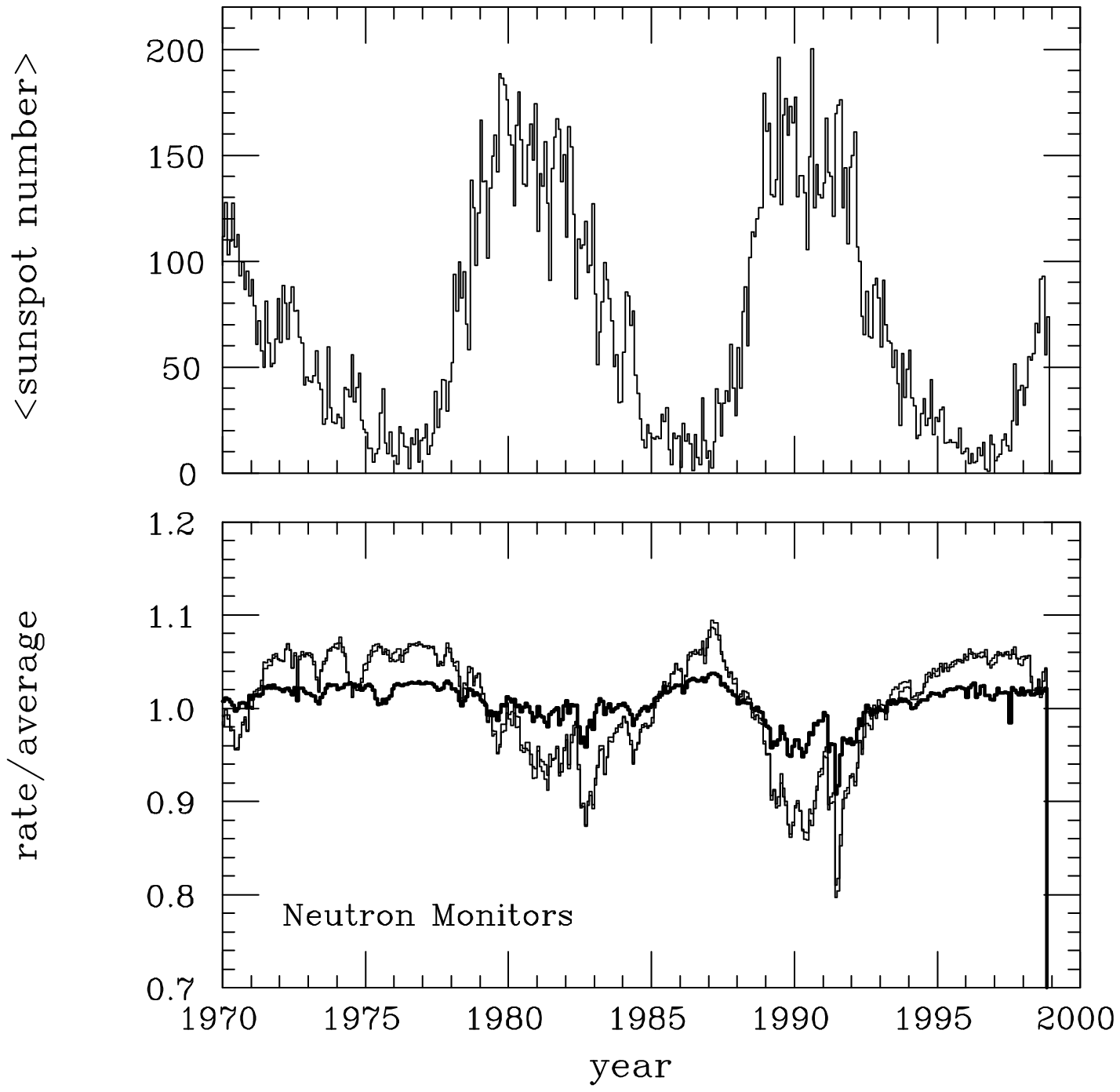,height=11cm}}
\fcaption {  The top  panel  shows the monthly  average of 
the Sunspot Number. The  bottom  banel  shows the 
relative rate of three neutron--monitor detectors,  
Kiel,  Moscow  and Huancayo (see text).
\label{fig:n_monitor}}
\end{figure}

The time  variability of the  primary
cosmic  ray  flux  is caused  by 
the  time  varying    structure  of the heliosphere,  that is 
on the  electro--magnetic  fields   that  fills  the interplanetary space 
and  are the result of  the interaction of the solar  wind
(an outflow  of plasma from the sun)
with the interplanetary medium.
The  topics  of the structure of the heliosphere, and the phenomenon
of the modulation  of cosmic rays
(the two topics  are of course strictly related  since
the c.r. modulations  are an important tool to study 
the characteristics and  dynamics  of the heliosphere)
are  very complex  (see \cite{Bieber} for a  review  and
references) and will not  be discussed  here.
It  can only    be said that because of the presence 
of the solar wind  plasma  and  the e.m.  fields associated with it,
for  cosmic rays   the
heliosphere   is  a  medium  that   goes  from semi--transparent 
(for  rigidity  $R \sim  5$~GeV)   to totally  opaque  (for    kinetic
energy of  few hundred MeV/nucleon). The transparency of the medium
depends on the intensity of  the  solar  wind, that is   correlated 
with the   intensity of the solar magnetic  activity,  
and varies with an    11 years  period   (22 years  considering also the
polarity of the  magnetic field). This   magnetic  activity 
can   be monitored for  example    with   measurements of
the number of sunspots.
The  correlation between the cosmic ray flux and the solar  activity
is shown in fig.~\ref{fig:n_monitor}.
The intensity of the cosmic rays at the Earth 
in the energy region 
$E_0 \sim 1$--20~GeV  can be  continuosly studied   at  ground level  
using neutron monitor  detectors.
These detectors measure the   hadronic  component  of 
secondary  cosmic  rays,   detecting 
the  neutrons  produced   in the interactions
of the secondary particles in the  material  (lead)  that 
surrounds the detector. The neutrons  are moderated 
(with paraffin or polyethylene)  and  are detected
as thermal particles using Boron tri--fluoride proportional counters.
The rate  measured  by a neutron--monitor detectors  depends
on the  magnetic  latitude of the detector  
that determines the minimum rigidity of the primary particles  that  can 
reach the    ground level, and by the  altitude  of the  detector,
that  determines  the absorption of the secondary particles.
The top panel of  fig.~\ref{fig:n_monitor}.
shows  the monthly average of the 
Sunspot Number   since 1970,  the bottom  panel  shows the variations
in the monthly counting rate of   three  neutron  monitors\cite{n-monitor}.
The two  thin histograms that  shows 
a  stronger  variability 
give the  activity  measured  by  two  detectors  located 
at Kiel  (altitude  54  meters,  vertical  cutoff 2.36 GV)  and
Moscow (altitude  200  meters,  vertical  cutoff 2.43 GV).
The two histograms  are  nearly superimposed  showing excellent 
correlation, they are measuring real variations  of the 
primary c.r. flux.
The thick  histogram   exhibiting a smaller variation is
the  activity  measured at  Huancayo  (Peru)
(altitude 3400~m,  cutoff  Rigidity 12.92~GV).
It can be seen   how the  time  variability of the cosmic  ray  flux 
is  reduced  to less that 5\% for this higher   rigidity cutoff.

\subsection {Geomagnetic effects}
The geomagnetic  field  has  some  very important effects
on  the  propagation of  cosmic  rays, and
forbids  low   rigidity  particles
from  reaching the surface of the Earth. 
The geomagnetic effects  depend  strongly  
on the position on the Earth  of the observer, and
on the direction of  observation.
For a qualitative  understanding   let us  approximate  the Earth
magnetic field  as a dipole, and  consider   two observers,
the first at one  magnetic pole, and   the second
on the magnetic   equator.
The `pole observer' looking  vertically  up,
can observe particles  of all   momenta,
since  in  this  case  the  field 
$\vec{B}$ and the  particle  momentum  $\vec{p}$ are   parallel 
and there is  no  bending
of the particle  trajectories;
the  observer   located   at the equator
when looking 
for particles  arriving  horizontally from east
can detect  protons  (or  particles  with  charge 
$+e$)   only if they have 
a momentum  larger  than approximately  59~GeV.
Particles with this   charge  and  momentum
have an (unstable)   circular trajectory
that  remains at  a constant radius $r = r_\oplus$   seeing a 
magnetic field  of  constant intensity ($|\vec{B}| \simeq  0.31$~Gauss)
always  orthogonal  to their 3-momentum. It is  obvious  that
 particles  with  lower  momentum  have   `impossible'  trajectories.
With   a little  more  work one   can   calculate  for example
that  the  `equatorial  observer'  looking  
horizontally west  (vertically  
up)  can observe protons  with    momentum    $p \aprge  10$~GeV
($p \aprge  15$~GeV).   In general   for  a detector position
and   direction  of  observation   ($\vec{x}$,$\Omega$), it is possible
to  compute a `cutoff  rigidity'  $R_c (\vec{x}, \Omega)$ 
 ($0 \aprle R_c \aprle 59$~GV) 
such  that  only particles  with   $R \ge R_c$  can be  measured.

The Liouville theorem  states  that the density of points  
in phase  space volume is  constant.   This  fundamental theorem
has a simple  and  important  application\cite{Lemaitre-Vallarta}
to the propagation of cosmic  rays in a static magnetic  field.
Since the  momentum of the particles  does  not  change in
absolute  value, then  the  differential  flux
along a  particle trajectory is constant, and  it  follows  that
if the  flux is isotropic  at large distances
from the Earth, then the flux in a small cone
around  any  trajectory   is  constant and  independent  from 
the position. More  formally 
if $\phi_\infty (p)$ is the isotropic flux  at large  distances  from the Earth,
then the flux    measured   by a detector  at a point 
$\vec{x}$  in the direction
$\Omega$    has the form:
\begin {equation}
\phi_{\vec{x}} (p, \Omega) = \phi_\infty  (p) \times
\zeta(p, \Omega, \vec{x})
\end{equation}
where the quantity $\zeta$ is either  0  or 1,  but never
something else.
The cosmic ray energy spectrum
is therefore not deformed by the  geomagnetic  field
and the effect  of $\vec{B}_\oplus$ is 
is to `remove'  a set of  momenta   from the spectrum.
In summary the   calculation of   the geomagnetic  effects,
using the facts that  field $\vec{B}$  is  approximately constant in 
time and the  primary  flux  is  isotropic at  large distances,
can  be  reduced  to the calculation  of  the allowed  and
forbidden trajectories.

This  problem can  be solved analytically  for  
the  special case of a volume    that  is  entirely filled 
with an  exactly  dipolar magnetic  field.
In this  case  the function  $\zeta$ has  the  form
of a single sharp step\cite{Stormer,Cooke}.  Introducing the 
rigidity $R = p/q$   the solution is:
\begin{equation}
\zeta (R, \Omega, \vec{x})  = \Theta [ R - R_S (r,\lambda, \theta, \varphi) ]
\label{eq:Stormer0}
\end{equation}
where $\Theta (x) $ is the  Heavyside function ($\Theta(x) = 0$ for 
$x < 1$, $\Theta(x) = 1$ for $ x \ge 1$),  $r$ is the distance from 
the dipole  center, $\lambda$ is the magnetic latitude, $\theta$ is 
the zenith angle, $\varphi$ is the azimuth angle   measured clockwise 
from the magnetic north, and $R_S$ is the St\"ormer\cite{Stormer}
rigidity cutoff 
that for positively charged  particles is:
\begin {equation}
R_S (r, \lambda, \theta, \varphi) =  \left ( {M \over 2 r^2 }
\right )  ~ \left \{  { \cos^4 \lambda \over
[1 + (1 - \cos^3 \lambda \sin \theta \sin \varphi)^{1/2}]^2 } \right \}
\label{eq:Stormer}
\end{equation}
(for  negatively  charged particles the cutoff
is obtained   with the   reflection 
$\varphi \to \varphi + \pi$)  that  is exchanging 
east  and west).
In (\ref{eq:Stormer}) $M$  is the magnetic dipole moment of  field.
For  the Earth
$M \simeq 8.1 \times 10^{25}$~Gauss~cm$^3$ that corresponds to a polar
magnetic  field  of 0.62 Gauss. The quantity $M/2 r_\oplus^2 
\simeq 59.4$~GV corresponds to the rigidity of a particle 
in a circular orbit 
of radius $r_\oplus$ in the earth's magnetic equatorial plane.

The St\"ormer analytic  result
is  only an  approximation for a realistic calculation
because  of two effects: (i)  the  Earth  field  is not  exactly dipolar,
 (ii)  the field   does not  fill the entire space,
and  therefore a  fraction of  the trajectories that   are 
`allowed'  according to the St\"ormer formula  intersects the  
 surface of the  Earth  and is  in fact forbidden.
The problem 
can be solved  in  general  numerically using the 
backtracking method. 
In this  method  to 
determine  if a cosmic ray  trajectory ($\vec{x}, \vec{p}, q$) is 
allowed or not, one integrates numerically the equation of motion
for  a particle  of momentum  
$-\vec{p}$  and charge $-q$,  starting from  the point $\vec{x}$
using a  detailed description of
the geomagnetic field,
this  of course corresponds  to 
the  calculation of the past--trajectory of the particle considered.
If the   trajectory   `goes to infinity' 
without  intersecting the  surface of the earth, the initial  
trajectory is allowed:
$\zeta (\vec{p}, \vec{x},q) =1$. If   the  trajectory
intersects the Earth, or   remains  confined  within a finite  radius
the  trajectory is     forbidden:
$\zeta (\vec{p}, \vec{x},q) =0$. 
We  note  that  in the   exact  calculation 
the function  $\zeta(R, \Omega)$ has  not  anymore  the form
of a  single  step  function,   and in a  narrow range  of   $p/q$
 the are alternating   intervals  of  allowed and  forbidden 
rigidities   (the `penumbra'  region), therefore 
strictly speaking 
 for a  fixed    position $\vec{x}$  and  direction $\Omega$ one 
should  not speak about a  single cutoff
rigidity  but  of  a  minimum allowed rigidity  $R_1$   and
a maximum  forbidden   rigidity $R_2$  (with $R_1  \le R_2$),
however since   $R_1 \simeq  R_2$  the concept of cutoff
remains  qualitatively  useful.
Both the Bartol\cite{lip-stan}  and HKKM\cite{HKKM} calculations 
use  the backtracking method to estimate the geomagnetic  effects.
An important  confirmation of the 
correctness of  the  estimate of the geomagnetic effects 
has  been obtained with the
experimental
observation of the  east-west anisotropy 
for atmospheric neutrinos  performed by SK\cite{SK-east-west}.

\subsection {Isotropy of the c.r.  flux}
In the previous  subsection we  have assumed  that the  
c.r. flux , in the absence of geomagnetic  effects   is isotropic.
This `intrinsic'  isotropy can  in principle be studied 
measuring the flux in  a fixed  direction
in the detector  system of coordinates  so that  the geomagnetic
effects are  constant.  Since the Earth is rotating
this  allows  to  study the variations of the 
primary  flux  in a cone in celestial   coordinates.
The  measurements   for the  range of  primary energy
of interest have  shown the  existence of 
 `solar' anisotropies (related  to  day--night  effects)
with   amplitudes of a fraction of a percent
and  no  `sidereal' anisotropies\cite{cr-anisotropies}.

\section {Modeling of the development of  hadronic  showers}
The  modeling of the development  of  hadronic  showers
is  a second  important source of uncertainty  in the calculation 
of the $\nu$ fluxes.
A cosmic ray shower   has a  `tree' structure
and can be seen as   made of  branches and    vertices.
The  primary particle interaction point    is
the   first vertex   of the tree,
the  secondary  particles produced in this
this interaction are a first set of branches.
Each one of the  secondary particles 
can be followed,  taking into account  energy loss,
deviations  in  the  geomagnetic  field and multiple scattering,
until  the  particle, decays,  interacts, stops, or  
reaches  the detector   level.
Decays and interactions  can be seen as   additional   vertices,
where  one  `in' particle    is destroyed  and
several `out'  particles  (or new  branches)  
are produced. The new particles  can also be followed,
and the process recursively   iterated.
The  neutrinos  are  produced in 
the   weak  decays   of  kaons and charged  pions.
Because of this  tree structure, the
calculation of    cosmic  ray  shower is  a problem
that can  be solved in the most  natural and 
accurate way with Montecarlo  techniques.
Analytic  treatments are also possible\cite{BN,lep-atm}
and represent a  reasonable approximation that allows to  gain insight 
on the physics of the problem,  however  the MC   methods
are more  general and  precise    and have to be preferred.

\subsection{Modeling of the hadronic  interactions}
The main uncertainty in the 
modeling of   hadronic  showers  is 
the description  of particle  production in
hadronic  interactions.
The multiplicity,  energy spectrum  and
flavor    composition of the particles produced in the final  state  
are all important factors  in the  determination
of the $\nu$ fluxes.

\begin{figure} [bt]
\centerline{\psfig{figure=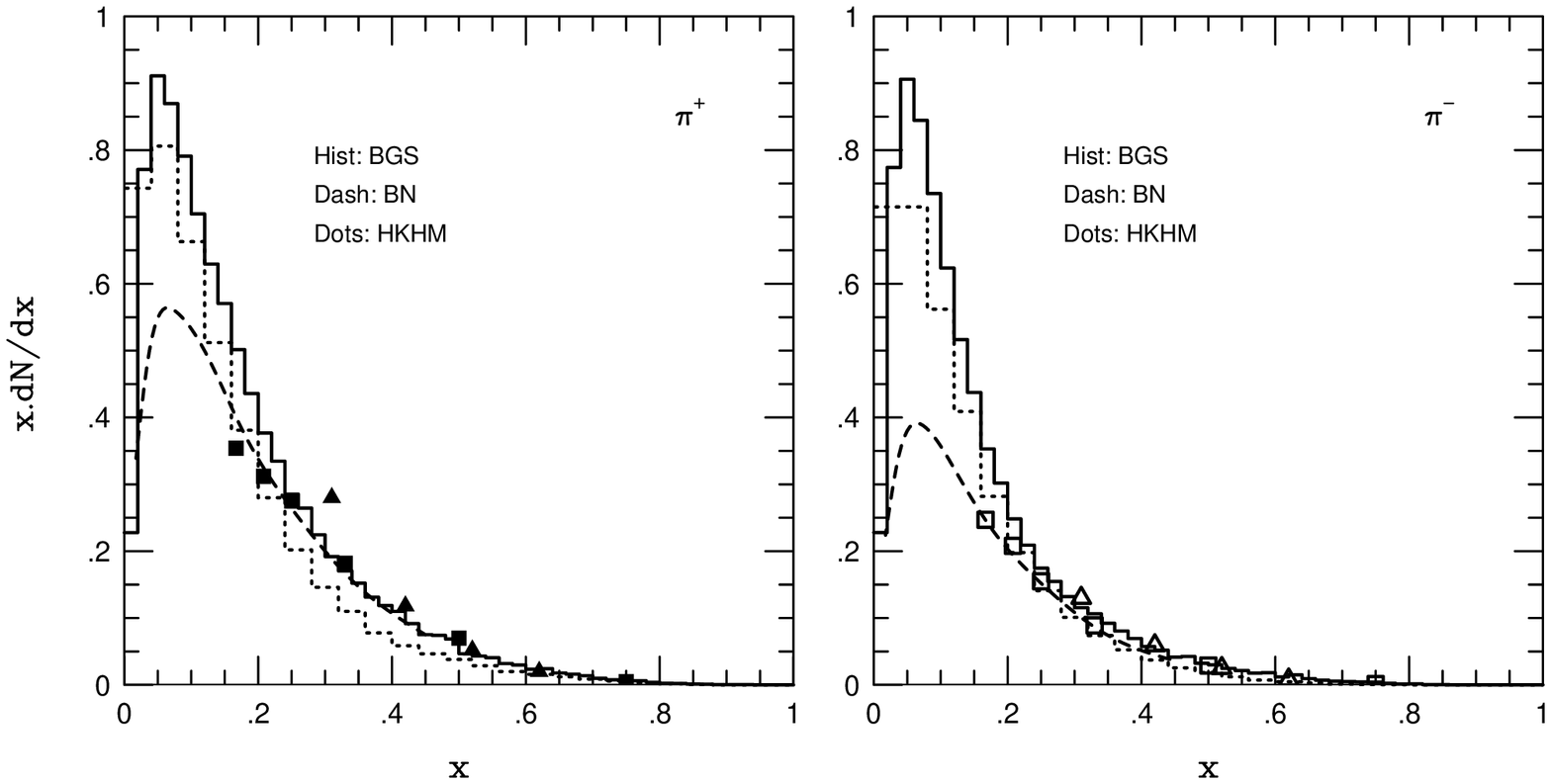,height=9cm,angle=90}}
\fcaption {From \protect\cite{atm-comp}.
Distributions of fractional momentum of $\pi^\pm$ 
produced in the interactions of $\simeq 20$~GeV$/c$
momentum protons with light nuclei  obtained   in three
different  models:
Bartol (BGS)\protect\cite{Bartol}, HKKM\protect\cite{HKKM} and
Bugaev and Naumov\protect\cite{BN}
for  hadronic  interactions on an air target.
The points are  estimates  obtained
is\protect\cite{atm-comp}  integrating on $p_\perp$ the 
data   on Beryllium\protect\cite{had-data}.
\label{fig:had}}
\end{figure}

In fig.~\ref{fig:had}   taken  from\cite{atm-comp} a    review  paper
on the calculation of atmospheric  $\nu$  fluxes
we show the energy distribution   of charged  pions 
produced in the  interactions  of  24~GeV protons  on 
air   as  modeled in  different  calculation of the
atmospheric  $\nu$ fluxes. 
Inspection of fig.~\ref{fig:had}  shows  that the Bartol   calculation
has  a higher multiplicity   and a slightly harder  energy spectrum
of the HKKM one,    and much  larger  difference  with respect
to the Bugaev and Naumov estimates.
This difference   in the modeling    of  particle  production
would  account   for a $\sim 20$\%   difference in  the  $\nu$ 
event  rate between the Bartol and HKKM  calculation if the same
primary flux  is assumed.

A  comparison   of  the  FLUKA\cite{fluka}  and Bartol\cite{Bartol} 
interaction models,   
shows  that   the FLUKA  code also predicts a smaller  multiplicity 
for charged  pions,   this  will result in a 
smaller $\nu$  flux    for the same 
primary flux. The  details  of the    full calculation
will soon be   made  available.

\subsection{The method of calculation:  3D versus 1D}
The MC calculation of   an  hadronic  shower
is   in principle a  straightforward,   even if
technically not  trivial, computational problem.
All the physics  that is  needed  is   well known, 
with the only exception of the description  of the hadronic  cross
sections, discussed in the  previous subsection.
However all the  calculations  that are 
  publically available 
(with the exception of a first attempt by Lee and Koh\cite{LK})
have been   made   in the approximation  of 
considering the   development  of the  hadronic showers  as one--dimensional.
This  implies (discussing  in particular the 
implementation of the approximation in the Bartol code) 
the following:
(a) multiple  scattering and  deviations in the magnetic  field
are  neglected: all  particles  propagate along  straight lines;
(b) in the  generation of the interaction or decay vertices
the  kinematical properties of  final state  particle  are
calculated exactly  including the  generation of 
$p_\perp$ and    the exact conservation of
4--momentum, however    after the   generation of the vertex  is  completed,
for  all particles with $p_\parallel > 0$  the
angle  with respect to  the primary particle  is  set to zero
while  particles with 
$p_\parallel < 0$  are  disregarded.
In this  way   each   $\nu$ is  exactly collinear with 
its primary particle.

The  motivations  for this  approximation  are  purely technical:
the saving  of computer time.    The neutrinos  that  reach a  detector
in  a certain geographycal position  (say  Super--Kamiokande)
are  in general produced  in  cosmic ray showers
that   developed  anywhere on the earth and   (allowing  for
$\nu$'s non collinear with the primary  particle)  with 
any  direction.
It is  clear  that the  Montecarlo generation in this general 
scheme is very inefficient,  because one  has  to 
generate    {\em all}   possible  cosmic  ray   showers
in the Earth  atmosphere,  but only  a 
very small  fraction of  them
will   produce  neutrinos in the vicinity of the detector.
With the assumption of collinearity  only a  very small  subset of 
the  cosmic ray    has  to be  studied  and the 
problem   becomes   numerically  enormously simpler.

The  dominant  contribution
to the angle $\theta_{0\nu}$ between the  neutrino  
and the parent particle arises  because  of the non--collinearity of the 
parent  mesons  with the primary particle,  and (for  $\nu$ from muon decay)
because of the deviation of the $\mu^\pm$ is the geomagnetic  field
with the  $\pi\nu$  angle a smaller  contribution:
\begin{equation}
\langle \theta_{0\nu} \rangle  
 \simeq 
\theta_{0\pi} 
\oplus \theta (\mu^\pm, B_{\oplus})  \simeq 
 {\langle p_{\perp \pi} \rangle \over p_\pi }
\oplus   \left ( {L_\mu   ~B_{\oplus \perp}   \over p_\mu }  \right ) 
\sim  {4.3^\circ \over E_\nu ({\rm GeV} )  }
\oplus   {10^\circ  \over  E_\nu ({\rm GeV})  }
\end{equation}

A first  test   of the  importance of  3D effects 
in  the calculation of  the atmospheric  $\nu$   fluxes 
has  been  performed   with FLUKA\cite{nu-fluka}. 
The test is  a  comparison   of   two  calculations of
the $\nu$--fluxes  performed   with the  1D approximation   or
 with a  fully   3D method, but in both cases 
assuming spherical  symmetry for the Earth and  
the  absence of the geomagnetic field.
In  this   approximation, all  points on the surface of  the Earth are 
equivalent and can be considered as a
single  detector, therefore  the CPU time  problem is  not present,
note however that since  the  deviation of the  muons  
in the  geomagnetic  field  is  neglected, 
the   angle  between $\nu$  and primary particle  is  understimated.
The  results  of the test show that
the  effects of the 1--dimensional  approximation
are  negligible for   multi--GeV events, and  a small 
correction   for  sub-GeV  events.
A complete  3D calculation is  feasible  using an efficient  weighting of
the simulated events.  Results on fully 3D 
calculations of the $\nu$  fluxes should very soon be made publically
available  (at least) from  the Fluka  and Bartol  groups.

\subsection {The  density of the atmosphere and  the mountains}
The density of  the medium where the c.r.  shower develops,
changes   with continuity  and it is essential
to take  this   fact into account.
The density profile of the atmosphere  can  to a good approximation
be taken  as constant in time  and  independent of  position
(the  concept of altimeter  is based on this approximation).
Variations  of the average density profile
with  the geographical latitude  and  the season  have  been investigated
and  are  the source only of minor  corrections for low
energy  neutrinos.
For  accurate  calculations  it is also  necessary to
consider the  profile  of the mountains above 
the detector. This  is  straightforward to  do, and   represents
a few percent reduction of the component of
high  energy  and  vertical  down--going neutrinos that is
produced in muon decay, the  effect  is  therefore 
stronger for electron neutrinos.

\section {The neutrino cross section}
The description  of the neutrino cross section  is a source of uncertainty
of comparable importance to  the primary  flux
and  the  hadronic cross sections   in the prediction
of the atmospheric  $\nu$  event rates.
At  high   $E_\nu$, when most of the   phase space
for  $\nu$ interactions  is in  the deep--inelastic   region,
$\sigma_\nu$ is   reliably calculable in  terms  of  well  determined
parton distribution  functions (PDF's).
However   in the energy range  relevant for
atmospheric  neutrinos ($E_\nu \sim 1$~GeV)
the  description of  $\sigma_\nu$ is  
theoretically   more difficult, and the  available  data is not sufficient
to constrain the cross  section to better than 15--20\%.

Quasi--elastic scattering  is
the  most  important  mode,  its  importance being  enhanced by 
the event  selection  criterion   (the `single ring' condition)
that  requires a single visible  particle in the final state.
Uncertainties on this process arise  from  uncertainties in the
axial  form factor  of the nucleons  (that  cannot be  measured in
electron  scattering), and  more important from the
description of the  nuclear effect  corrections.
Most MC codes use  some  version of the  Fermi gas model
\cite{Smith-Moniz}.  More   sophisticated studies 
\cite{Engel,Marteau}  have  attempted  to   calculate
the nuclear effects  beyond the  Fermi gas  moded  using different
theoretical  methods to describe the  nuclear  medium.
 These  studies  were originally  developed 
to see if nuclear  effects could  somehow result in 
different  cross  section for $\nu_e$'s  and $\nu_\mu$'s,
and so explain the low  $(\mu /e$) ratio  observed  experimentally.
No such  flavor  dependence  has  been  found,
however the   calculations 
have  also estimated 
significant effects  (10--15\%)  in the absolute value of 
the QE cross sections.
The Soudan detector  in a good  fraction 
of the events  can  measure  together with the charged  lepton, 
also the  recoil proton 
in processes as:  $\nu_\mu \,Fe \to \mu^- p \,A^*$,
where  $A^*$ is  an undetected nuclear system.
The resolutions
that can  be attained    for the  neutrino 
kinematical  properties ($E_\nu$ and $\Omega_\nu$)  depend
on  how  much 4--momentum is carried away fron the undetected 
nuclear system  (the `spectator'  in the simple Fermi gas model), and
can   depend on the modeling  of the  nuclear effects.
A detailed study  should    investigate   if
the resolutions  estimated in the Fermi gas  model\cite{Soudan}
are  correct.

\begin{figure} [bt]
\begin {minipage} [h] {6.0in}
\psfig{figure=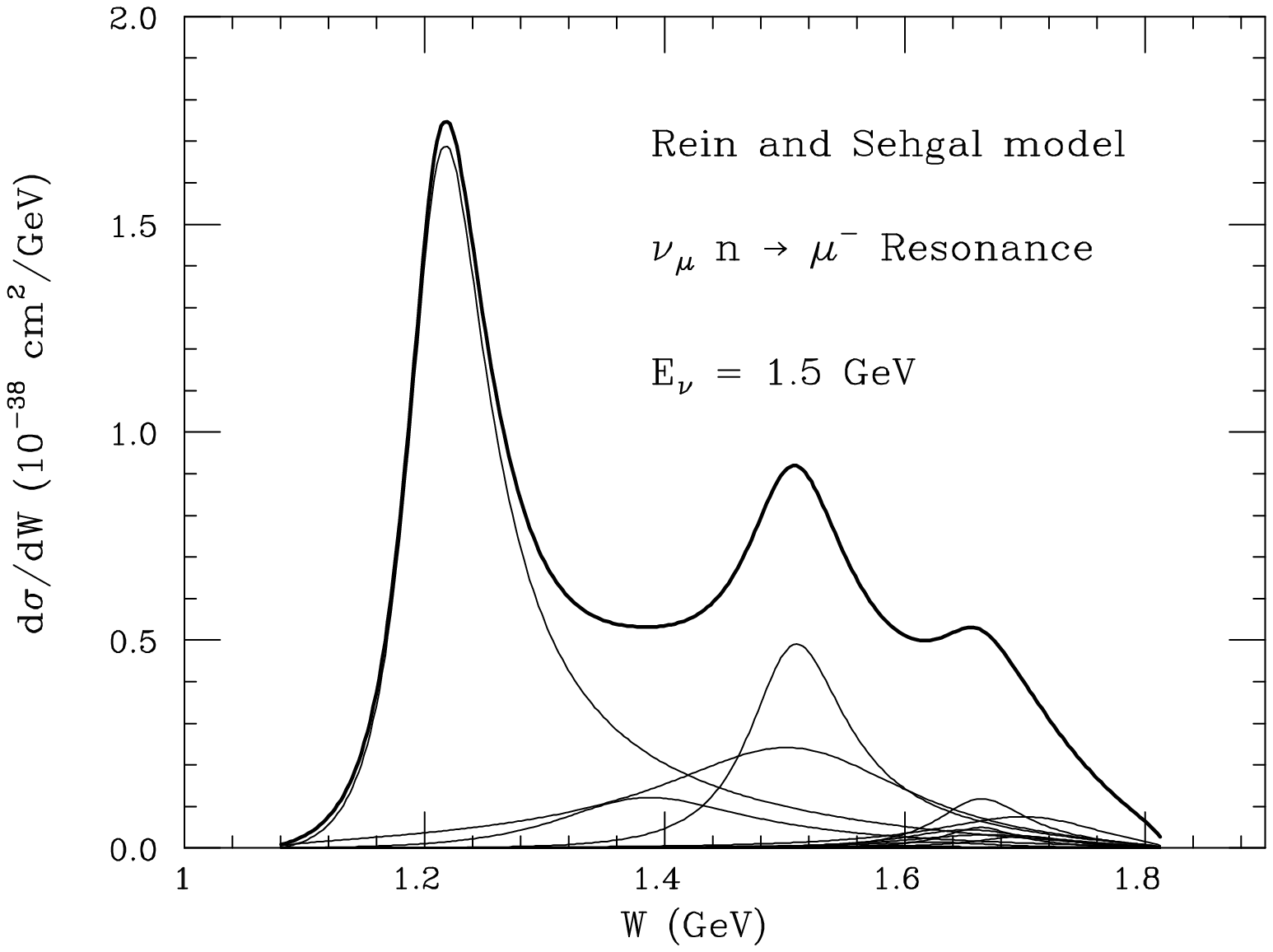,height=5.6cm}
\psfig{figure=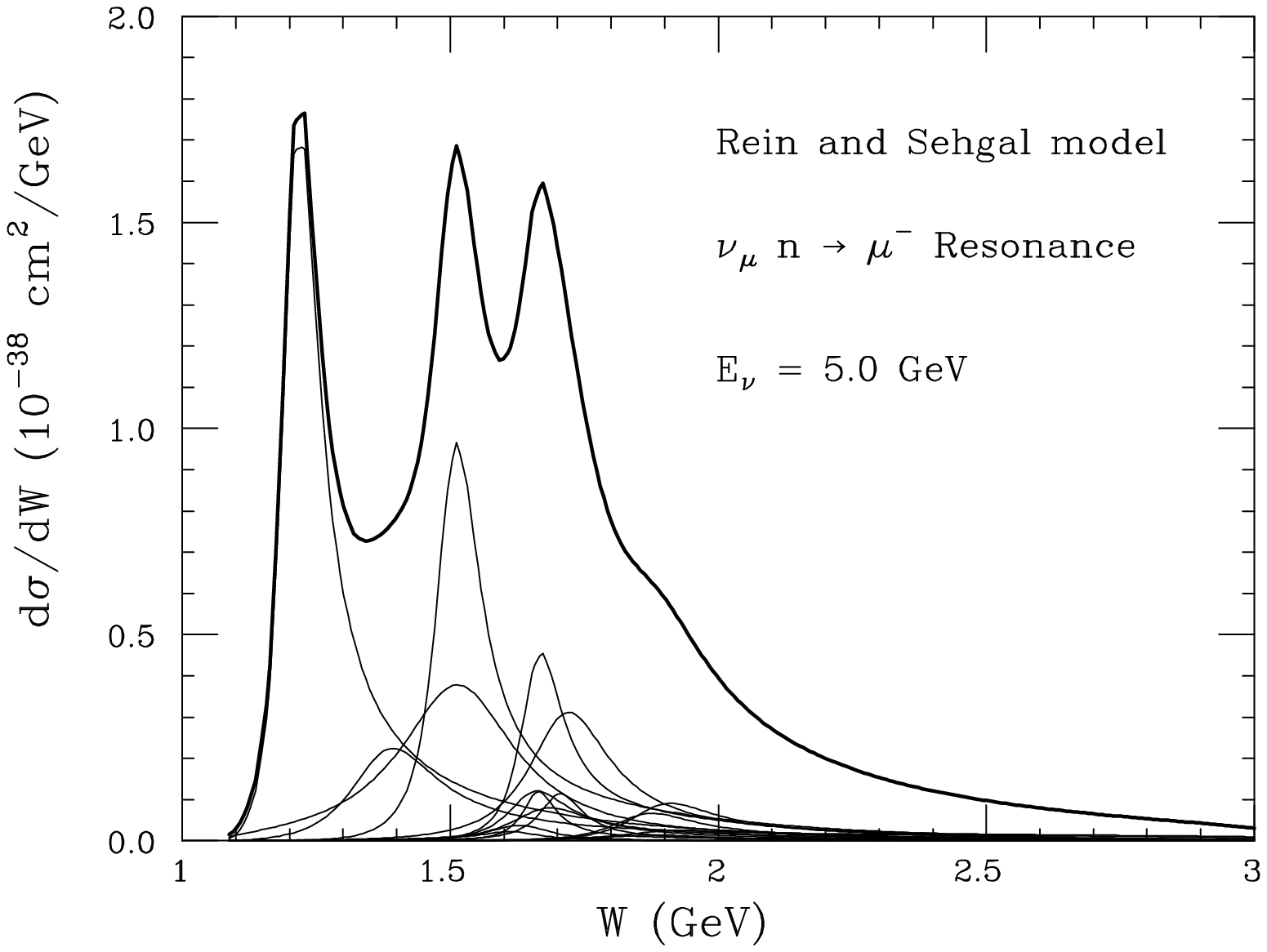,height=5.6cm}
\end{minipage}
\fcaption {Differential  cross section
$d\sigma/dW$ for the  scattering
$\nu_\mu  n \to \mu^- R$   where  $R$ is  a  baryonic  resonance,
calculated  according to\protect\cite{Rein-Sehgal}.
The thick  curve is the sum of all contributions.
\label{fig:rs}}
\end{figure}

Processes  where the   target nucleon is  excited to 
to a resonance  are  also  of  great importance in the
energy region  relevant for atmospheric neutrinos.
Both the Super--Kamiokande and Soudan   MC codes  describe
this cross section using the  algorithms of Rein and Sehgal\cite{Rein-Sehgal}
that apply to neutrino  interactions  the  model of 
baryonic  resonances of Feynman, Kislinger and Ravndal\cite{FKR}
that  describes  the baryons  as   bound  states   of three valence
quark in  an harmonic oscillator potential.
Two examples of the prediction
for $d\sigma/dW$   ($W$  is the mass of the    hadronic
system in the final state), calculated according to the
original  work,
are shown as an  illustration in fig.~\ref{fig:rs}.
We note that the Rein and Sehgal model  predicts  
a   complex  structure   in the   $W$   distribution,  and  at intermediate
energy not  only the  $\Delta$  resonance (the strongest peak
at $W \simeq 1.232$~GeV)  but also other states   give important  contribution
to the cross section.  The existing data does  not  give  a  clear
confirmation  of the  prediction.
Nuclear  effects are also  important  for  resonance production.
In particular   in  a  nuclear  medium there is the 
possibility of  a  baryonic   resonance  
that  is  dis--excited   to a nucleon, with  the scattering
appearing   experimentally as  QE  scattering. 

\begin{figure} [bt]
\begin {minipage} [h] {6.0in}
\psfig{figure=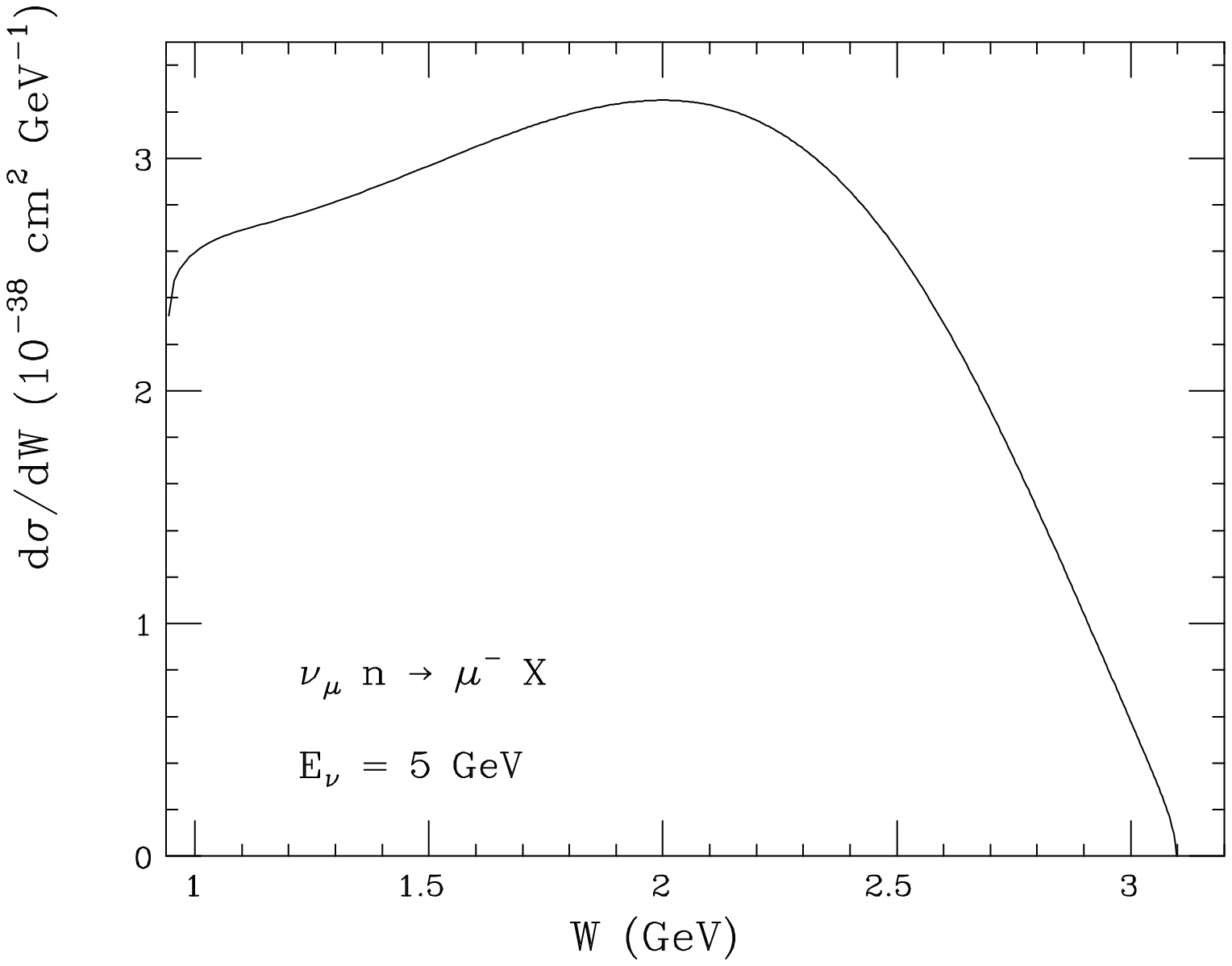,height=5.7cm}
\psfig{figure=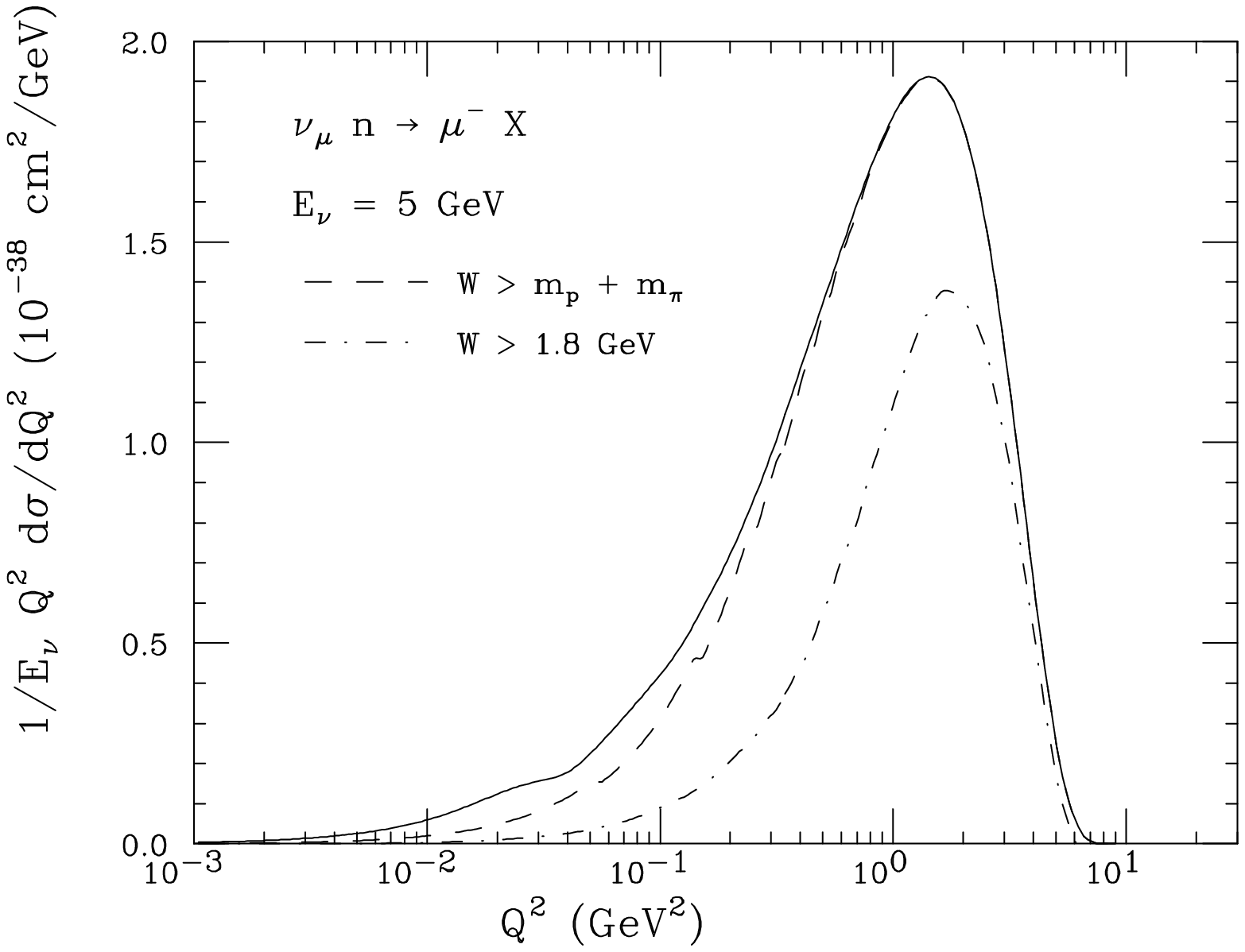,height=5.7cm}
\end{minipage}
\fcaption {Left:  plot of $d\sigma/dW$ for  $\nu_\mu n \to n X$ 
interactions. Right:  plot of  $d\sigma/d\ln Q^2$   for 
the same  reaction. 
The  $\nu$ cross section is calculated  using the  DIS  formulae and
the GRV94LO PDF's. \label{fig:dis}}
\end{figure}

All other channels
in the $\nu$ cross section  are usually described
using the standard   expression for deep inelastic  scattering
(DIS)\cite{Albright-Jarlskog}.
In fig.~\ref{fig:dis}  we show and  examples of
the  differential cross sections 
$d\sigma/dW$ and  $d\sigma/d\ln Q^2$   calculated
using the  DIS  formulae, and the 
leading order PDF's  of GRV94\cite{grv}.
Looking at $d\sigma/dW$  in the left panel of
fig.~\ref{fig:dis}  one  can see that the 
as  expected the  DIS formula  does not  `know'  anything 
about  the hadronic mass spectrum, so  it does not  describe  resonances
and in fact   it computes   a cross section  also
in the  unphysical region $ m_p < W < (m_p + m_\pi)$   that is 
 forbidden  because no  hadronic state with   mass in this  interval
can exist,
It remains  an open problem  how to `join' smoothly the 
baryon resonance production  with the   all  other  channels  
described collectively by  DIS.
It should also  be  noted, that 
because of  the low  energy of atmospheric  $\nu$'s
the $Q^2$   in the  interactions is  always small,
even if  the  final  state hadronic  system is excited  above the
resonance  region (see the right panel
of fig.~\ref{fig:dis})   and therefore  the use of the  DIS  framework is
not on a solid ground.

The importance of the neutrino  cross section in the prediction of the
atmospheric neutrino event rates    is  revealed by
the effect of a   relatively small modification  of the
description of $\sigma_\nu$ in the SK  Montecarlo:
the choice of  a  new set  of  PDF's
(GRV94LO\cite{grv}  replacing  the CCFR  parametrization)
to describe the DIS part of  $\sigma_\nu$.
Combined  with the  use  of the   medium solar
activity $\nu$ fluxes of HKKM\cite{HKKM}  (that  has a small impact
for  multi--GeV events)  has led for example to  an increase
of 4\% (8\%) for the  $\mu$--like  ($e$--like) fully contained  events, 
and  7\% for the partially contained  events
(compare the  MC predictions  in\cite{SK} and\cite{SK-new}).

It appears  very  difficult  to  calculate  accurately 
$\sigma_\nu$ from first principles in the 
relevant  energy region.  The existing  data     do  not 
determine  the   absolute  value of  $\sigma_\nu$
and the energy spectrum of the final state lepton better that 
$\sim 15\%$.   A significant  improvement  can probably only be  obtained
with   the  analysis of more data.
The  K2K long--baseline $\nu$ beam\cite{K2K} 
 with a spectrum not too different
from the  atmospheric one offers  very interesting  possibilities.
The  placement in the `near'  position
along the beam  of detectors 
with higher  resolution  and/or an   iron  or  argon target
could  also  be useful for   a more complete   study.

One  final  remark on the  question of  the $\nu$ cross section,
is that   all the experimental groups
that have  made   measurements of  atmospheric  $\nu$'s have 
compared  their results to  predictions  calculated 
using different  descriptions of  $\sigma_\nu$  and
different  MC codes (in general
not publically available) for the  generation of  the $\nu$ interactions.
A  detailed  comparison  of the  descriptions of $\sigma_\nu$ used
by the  different  groups is  missing.
This  lack of information   can be a  source of  confusion.
For example  it is important  to understand 
the   reasons for  the difference in the 
ratios  $e_{\rm data}/e_{\rm MC}$  stated by the 
SK and Soudan  (see section 8) and establish it is  due
to the  $\nu$ flux or,
at least  in part, to 
different  assumptions  about the $\nu$cross sections.

\section{The muon flux constraints}

The fluxes  of atmospheric muons  are  strictly related
to  the neutrino ones,    because  almost all  $\nu$'s    
are   produced  either in association with, or in the decay of $\mu^\pm$.
Precise measurements  of the muons  offer then a way to check
directly the  $\nu$  fluxes.
To  constraint the flux   of neutrinos  with
energy $E_\nu$, the most interesting  muons
are   those created  with  energy  around 
$E_\mu \sim 2\; E_\nu$  (the  factor   of two being a
kinematical effect).  The most 
important range of   $\nu$   energy is $E_\nu \sim 0.1$--10~GeV,
and  especially for the lowest part of this range,
only a vanishing  or small  fraction of  the  related  muons 
reach the ground  because of   energy
loss  (the  atmosphere  corresponds to a loss
of $\sim  2$~GeV for a  vertical muon)  and decay
(the  $\mu$ decay length is $\ell_\mu \simeq 6.2~p_\mu($GeV)~km).
For  low  energy muons    the measurement 
is  possible  only at  high altitude.

\begin{figure} [bt]
\centerline{\psfig{figure=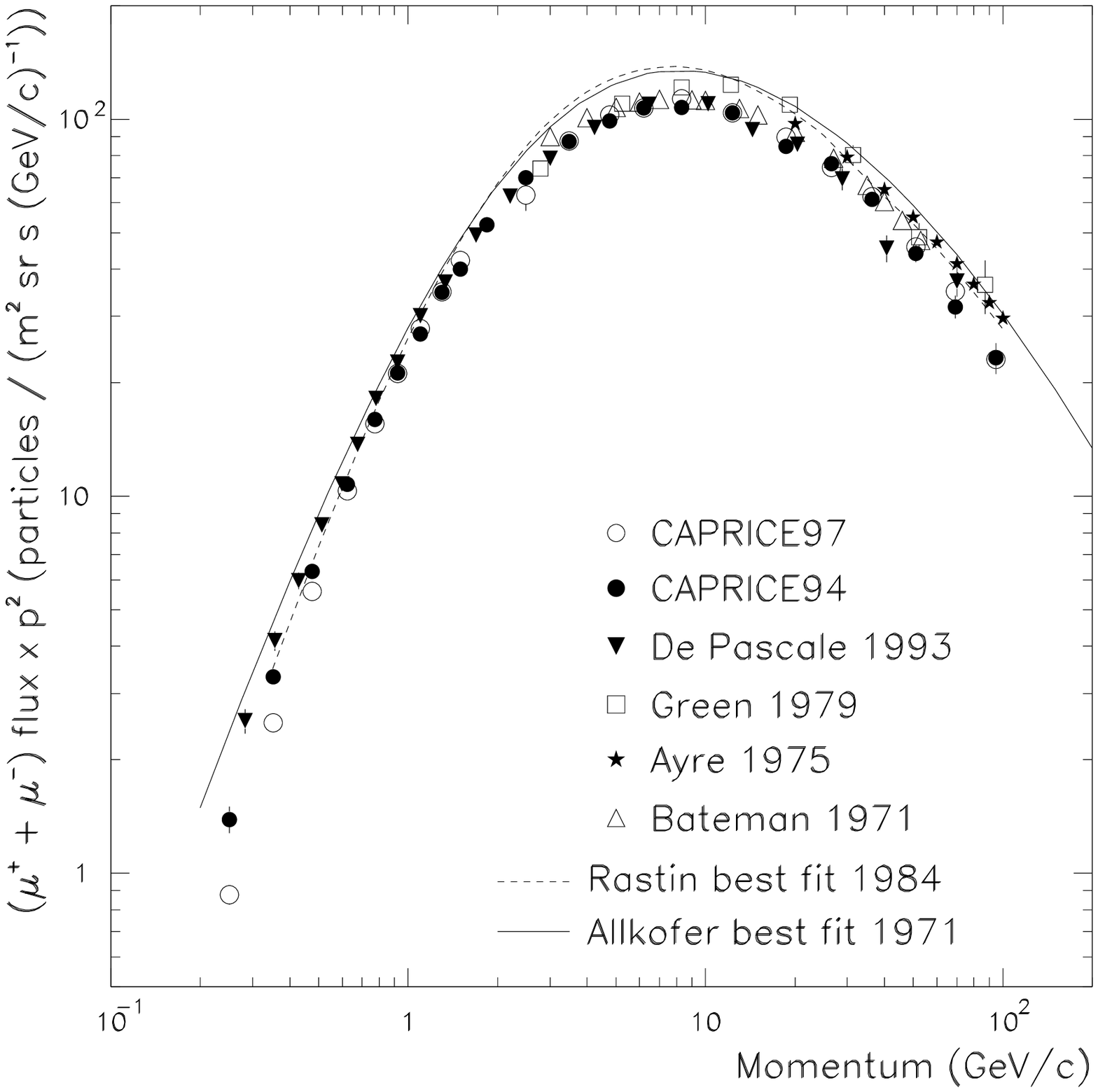,height=9cm}}
\fcaption {
Flux of  muons at the ground  measured by the 
Caprice detector\protect\cite{caprice-mu}, compared with  fit
to  previous  data.
\label{fig:caprice-mu}}
\end{figure}

The authors of  the existing atmospheric  $\nu$ calculations  
checked that their predictions  for  the  muon  flux (that is  `automatically'
calculated   together with the $\nu$ flux) with 
the existing muon data\cite{allkofer,rastin} at ground level.
To a large  extent  the agreement   
between the HKKM and Bartol calculations
of the  $\nu$ fluxes, that
is the effect of a cancellation  between   different  estimates
of the primary flux  and  different modelings of the 
 hadronic  interaction, is not    casual, but 
the  result  of the $\mu$ constraint.
Recently  new measurements\cite{mass-mu,caprice-mu,BESS}  of the  
ground  level  muon flux, have   been performed  with  detectors
developed for balloon  flights.
The new data    suggests    that 
previous  estimates  could be too high, 
(see for example fig.~\ref{fig:caprice-mu}).
This is  an important point that need to  be  studied  in detail.
New precision measurements  of muons  at  ground level
in sites with different  geomagnetic  cutoffs and at different altitudes
would  certainly be very valuable.

In recent years,  the balloon detectors have  performed  measurements
of the muon spectrum during their ascent  in the 
atmosphere\cite{MASS1,MASS2-mu,IMAXmu,HEAT}.
The  analysis  and  discussion of these  results  is  still in progress.
Preliminary comparisons  of these  results with MC calculations
have  shown some  significant discrepancies (especially  
for  $E_\mu \aprle 1$~GeV) that  require a  more  detailed study.
These are  difficult  measurements,   with the altitude of the
detector  continuosly  varying, 
and the  statistical errors are  still large  because of the 
short time  available  during   balloon ascent,
some  discrepancies  between the results of different  groups  
do exist, however these  and  future  measurements at  high altitude
(with detectors on balloons  or  possibly airplanes\cite{spy-sky})
have a good potential   for
reducing the uncertainties in the flux of atmospheric neutrinos.

\section {The `Normalization problem'}
In the  analysis of  their  data  the SK  and Soudan  collaboration
consider the absolute normalization of the   calculated
$\nu$  fluxes   as a  free parameter.
In the  SK analysis of the data  in term of   
$\nu_\mu \leftrightarrow \nu_\tau$ oscillations, the
$\nu$   fluxes are  fitted to  the form
\begin {eqnarray} 
\phi_{\nu_\mu} (E_\nu, \theta_z)  & = &
 (1 + \alpha) ~\phi_{\nu_\mu}^{{\rm th},0} (E_\nu, \theta_z) 
  ~ \langle
P_{\nu_\mu \to \nu_\mu}  
(E_\nu, \theta_z; \;\sin^2 2 \theta, \Delta m^2) \rangle \\
\phi_{\nu_e} (E_\nu, \theta_z)  & = &
 (1 + \alpha)~ \phi_{\nu_e}^{{\rm th},0} (E_\nu, \theta_z)  
\nonumber
\end{eqnarray}
where  $\phi_{\nu_\mu(\nu_e)}^{{\rm th}, 0}$  are the
published   no--oscillation fluxes   of
HKKM (or Bartol) and $\alpha$  is   an unconstrained  normalization
parameter.
In the analysis  of   535 days  live days of SK, 
for the best fit point,  the normalization parameter
had the value  $\alpha = +0.158$, 
also  the ratios  $e_{\rm data}/e_{\rm MC}$     
for the sub--GeV and  multi--GeV   samples.
had the  values
$1.18 \pm 0.03$ and $1.23 \pm 0.07$ (statistical errors  only).
In  other words    according to the analysis in\cite{SK}   the 
the  calculated   $\nu$  fluxes  of Honda et al.\cite{HKKM}
are too  low  by $\sim 15$--20\%.
This   result is  within  the stated   uncertainty in the 
absolute normalization of the calculation, however since 
the primary c.r. flux  used in the  HKKM  calculation
is  likely to be  too large,  this relatively high  measured  neutrino
rate has generated   discussions  and  speculations, at least in part
motivated by  the observation,   that
under the  hypothesis  of $\nu_\mu \leftrightarrow \nu_\tau$  oscillations
the  flux  if electron  neutrinos  is     not  affected  by oscillations,
however is    $\nu_\mu \leftrightarrow  \nu_e$   transitions  exist, 
it can be enhanced.

In the analysis  of   735 days  presented  at this conference
the  SK  collaboration estimates a lower  normalization   parameter
$\alpha = 0.084$.   This can  also be  seen  observing that the
ratios  $e_{\rm data}/e_{\rm MC}$  take the  new values
$1.06 \pm 0.03$ and $1.08 \pm 0.05$ (for  sub--GeV and multi--GeV).
The significant  differences with respect to the previous  analysis
are  due to a combination of  changes in the
MC  predictions and  of  statistical fluctuations 
in  the data\footnote{
The difference in the  estimates
of the rate of  sub--GeV  $e$--like   events  
for the  535 and  736  days  analysis
has  generated  
speculations\protect\cite{losecco}
that  the   $\nu$   fluxes   have  a time  varying component.
The analysis  in\protect\cite{losecco} does not take into account 
small  modifications  of the  pattern  reconstruction program that
affect  the  entire   integrated data sample. 
Taking this into account, the  evidence 
for a  transient   component  in   the $\nu$ flux becomes  much  weaker.}.
The  MC prediction has   increased 
because of  two  effects:
(i) the  data  is compared  to 
the $\nu$ fluxes    of  HKKM   calculated  
for  minimum   solar activity, while  in the    545  days analysis\cite{SK} 
the   data  was  compared  to  medium solar  activity\footnote{
The  use  of the solar minimum  fluxes is  certainly well motivated 
considering the  solar activity during the data taking  period
(see fig.~\ref{fig:n_monitor}),   the  higher  $p$ flux 
implied  by this assumption  could  however be in worst agreement 
with the  most recent data (see  fig.~\ref{fig:proton})
on the primary c.r. flux.};
(ii)   the description of the
$\nu$ cross section has  been  modified
as  discussed in  section 6 with a resulting 
enhancement of the event rate.
The combined  effects  of these changes are  an   enhancement of
the  MC rate of 4\% (8\%)   for the sub (multi)--GeV  sample.

\vspace{0.25 cm}
The   result of the  Soudan--2 experiment\cite{Soudan}
for the `shower' events is 
$e_{\rm data}/e_{\rm MC} = 0.81 \pm 0.09$, 
for  all   contained  events, 
($0.86 \pm 0.11$  for the  `high  resolution sample')
that is  an  effect  of opposite sign with respect to SK.
In this  case the   MC  calculation is 
based on the Bartol\cite{Bartol} 
$\nu$ fluxes.
The  difference 
in $e_{\rm data}/e_{\rm MC}$  between  SK and  Soudan
is  a  hint of a possible problem, and should  be 
carefully investigated. 
The  neutrino  flux   at Soudan is softer that at Kamioka
(see fig.~\ref{fig:nu_rate})  and is produced   on average by
lower  energy primaries  (see fig.~\ref{fig:response}),
therefore   the difference
between   the 
estimated ratios $e_{\rm data}/e_{\rm MC}$  at  Kamioka  and Soudan
is   not necessarily an experimental discrepancy,
but it could  be for example an indication  that the  calculated 
$\nu$ flux is too soft. In section 8,   we  will
discuss the possible effects of   a  modification  of the shape
 of the $\nu$  energy spectra.

\section {Determination of the oscillation parameters}
In this section I will  try to illustrate (only qualitatively)
possible  effects  of  the uncertainties in the predictions
of the  $\nu$ event rates  in the 
the determination of the oscillation parameters
in the hypothesis  of the  existence 
of $\nu_\mu \leftrightarrow \nu_\tau$ oscillations.

\subsection {The Soudan experiment} 
The Soudan experime\cite{Soudan}  has measured
for the `high resolution sample' a double ratio $R= 0.59 \pm 0.09$.
This  value is obtained as:  $0.59 =
0.74/1.25$. It is important to note that
the denominator  of the double ratio  $(\mu /e)_{\rm MC}$ 
is significantly smaller that the naive expectation of
approximately two\footnote{The same argument is also valid for the
single track/shower events, where the double ratio is $R = 0.66 \pm 0.11$, 
with $0.66 = 0.70/0.95$}. The  low  value 
of  $(\mu /e)_{\rm MC}$   can  be   easily   understood  as the
consequence  of  different detector efficiencies for
$\mu$--like and $e$--like events.  Muons are more penetrating
than electrons, and the containement requirement results in a 
reduced  efficiency for higher energy muons, and therefore in a low value for
for the $\mu/e$ ratio.  This implies that the ($\mu/e$)  ratio  is sensitive
to the {\em shape} of the $\nu$ energy spectrum with a  softer (harder)
spectrum resulting in a higher (lower)   ratio.
If  the  calculated  $\nu$  flux  (as an example)
falls  more  steeply with energy than the true  spectrum, 
 the denominator  of the double ratio is overestimated,
therefore  $R$ is  underestimated 
and   the estimate of the  oscillation
parameters is  biased: for a  fixed  value of $\sin^2 2 \theta$,
the  range of  allowed
$|\Delta m^2|$ is  overestimated.

As discussed in section 7, there is some experimental indication 
that  the shape of the $\nu$ energy spectrum is not 
predicted correctly  because the   measured 
event rate for sub--GeV  $e$--like events in 
Soudan  is  15--20\%  lower than the prediction,
while in Super--Kamiokande, where $\langle E_\nu\rangle$  is higher
because  of different  geomagnetic effects,  the detected
rate  is   6--8\% higher  than the prediction.
If this is  interpreted as   evidence  that the  $\nu$  spectrum is
flatter than the    prediction,   
the allowed  region for  $|\Delta m^2|$ estimated by Soudan
should be  revised    to  lower  values.

The Soudan experiment, differently from SK, has {\em not} observed a
  clear zenith angle modulation of the $\mu$--like event rate.  This
  can be interpreted as an indication that the $\nu$ oscillation length is
  sufficiently short, so that also down--going $\nu_\mu$ are
  significantly suppressed by oscillations, and can be used to exclude
  low values of $\Delta m^2$.  This conclusion is also  somewhat  model
  dependent, because 
  for the low energy $\nu$'s  detected by Soudan, the geomagnetic effects
  induce a non negligible up--down asymmetry, with an excess of
  up--down events.  The `geomagnetic asymmetry' has to be subtracted
  from the data 
  in order to put in evidence the possible suppression of up--going
  neutrino induced by oscillations.  From table~2 we can see for
  example that the Bartol prediction for the `geomagnetic asymmetry'
  at Soudan depends on the solar cycle epoch and decreases for the
  flatter spectrum of solar maximum activity.  Therefore a
  flatter neutrino energy spectrum\cite{barbieri}
   would predict of
  a smaller asymmetry  induced 
  by geomagnetic effect, then  the experimentally detected  asymmetry
  (that  is of  order $A = -0.25 \pm 0.16$) 
   would   be interpreted as more significant,
   and again  the allowed   interval  for
  $|\Delta m^2|$  for the $\nu_\mu \leftrightarrow
  \nu_\tau$ hypothesis  would be   shifted  to lower values.

\subsection {The Super--Kamiokande experiment}
In the case of the Super--Kamiokande experiment, the up--down
asymmetry represents really a `smoking gun' evidence for new physics.
The theory predicts unambiguously that at Kamioka the up--down
asymmetry is {\em positive} (excess of up--going particles) and of
order $A \simeq 0.01$ for the multi--GeV events.  The detected
asymmetry becomes more pronounced with increasing energy, precisely as
expected considering the improving correlation between neutrino and
muon direction, and the shape of the distortions of the zenith
angle distributions also match very well the expectations.  As a
consequence, the possibilities of a very large difference in
acceptance for up--going and down--going particles, or of the presence
of a very large non--neutrino background appear extremely contrived
and unlikely.

The up--down asymmetry of multi--GeV events is an excellent 
method to measure the mixing parameter $\sin^2 2 \theta$.
Since the pathlength for neutrinos coming from vertically up or down
differ by three orders of magnitude, it can be expected that for a
broad range of $|\Delta m^2|$,
if the correlation between the  $\nu_\mu$ and
the detected  $\mu$   directions is good, 
one will have that oscillations are absent for down--going
particles and can be averaged for up--going particles.  Selecting two
cones around the vertical, it is natural to expect that $U \simeq U_0
\,(1 - \sin^2 2 \theta/2)$, and $D \simeq D_0 \simeq U_0$. 
The mixing parameter can then be determined independently from the squared
mass difference as $\sin^2 2 \theta = 2 (1 - U/D)$  with  a
precision  that is essentially limited by the statistical errors.  Using
this `algorithm' the SK result for the asymmetry for the multi--GeV
data: $A = -0.31 \pm 0.04$ can be translated in a measurement of the
mixing parameter $\sin^2 2 \theta = -4A/(1-A) = 0.95 \pm 0.09$, a
result in good agreement with what is obtained in the detailed fit.

\begin{figure} [bt]
\centerline{\psfig{figure=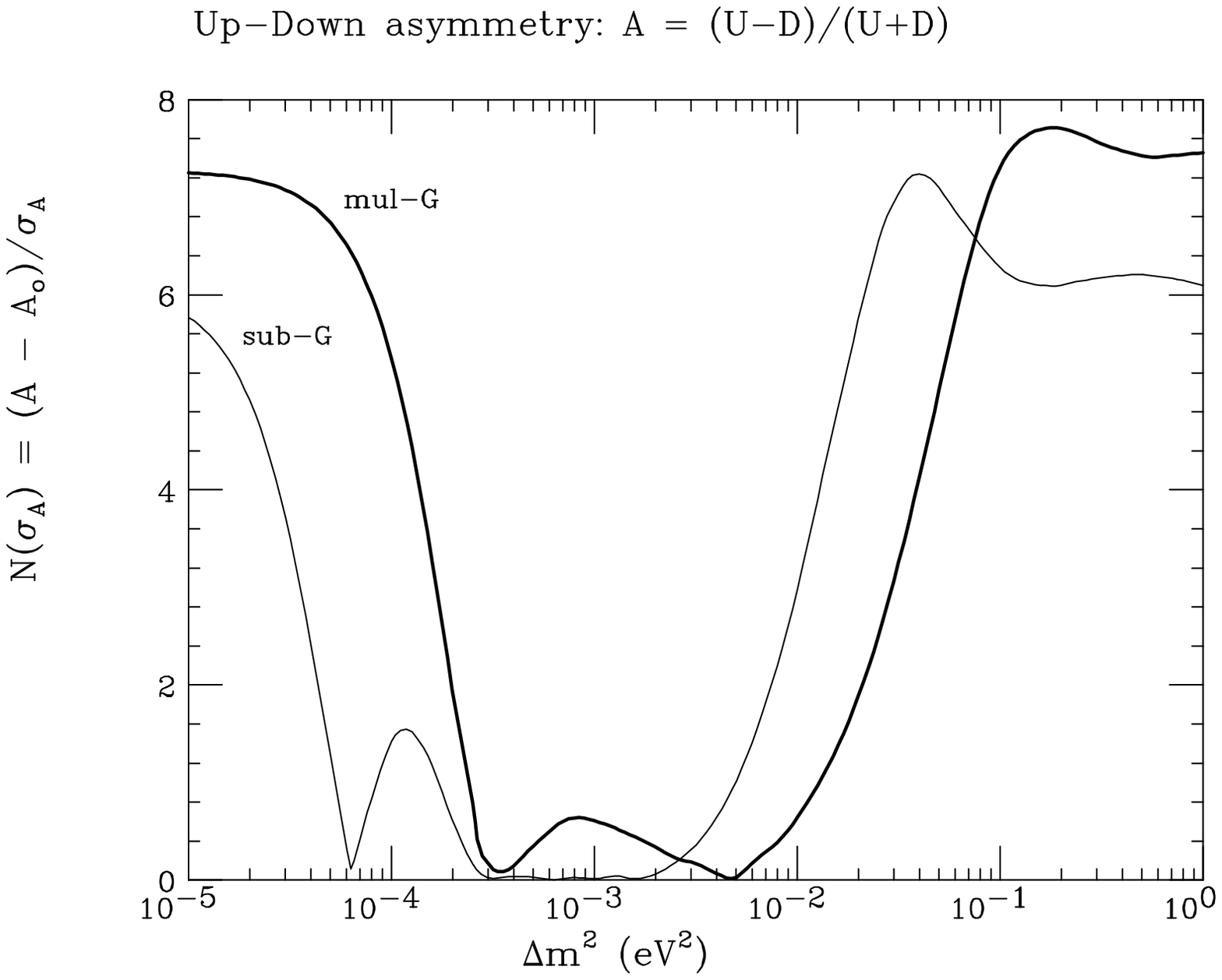,height=8.6cm}}
\fcaption {Number of  standard  deviations    for the
Up/Down  asymmetry  measured by SK. The two curves are for  sub--GeV
and multi--GeV data.
\label{fig:asym}}
\end{figure}

This argument however tells us that the up--down asymmetry is not a
good quantity for the measurement of $\Delta m^2$.  This is
illustrated in fig.~\ref{fig:asym} 
that shows as a function of
$|\Delta m^2|$ the number of standard deviations of the asymmetry
measurement for the sub--GeV and multi--GeV sample from
predictions calculated assuming the presence of $\nu_\mu
\leftrightarrow \nu_\tau$ oscillations with maximal mixing.  The
shapes of the curves show that there is a broad interval of $|\Delta
m^2|$ where the asymmetries have the measured value.  

\begin{figure} [bt]
\centerline{\psfig{figure=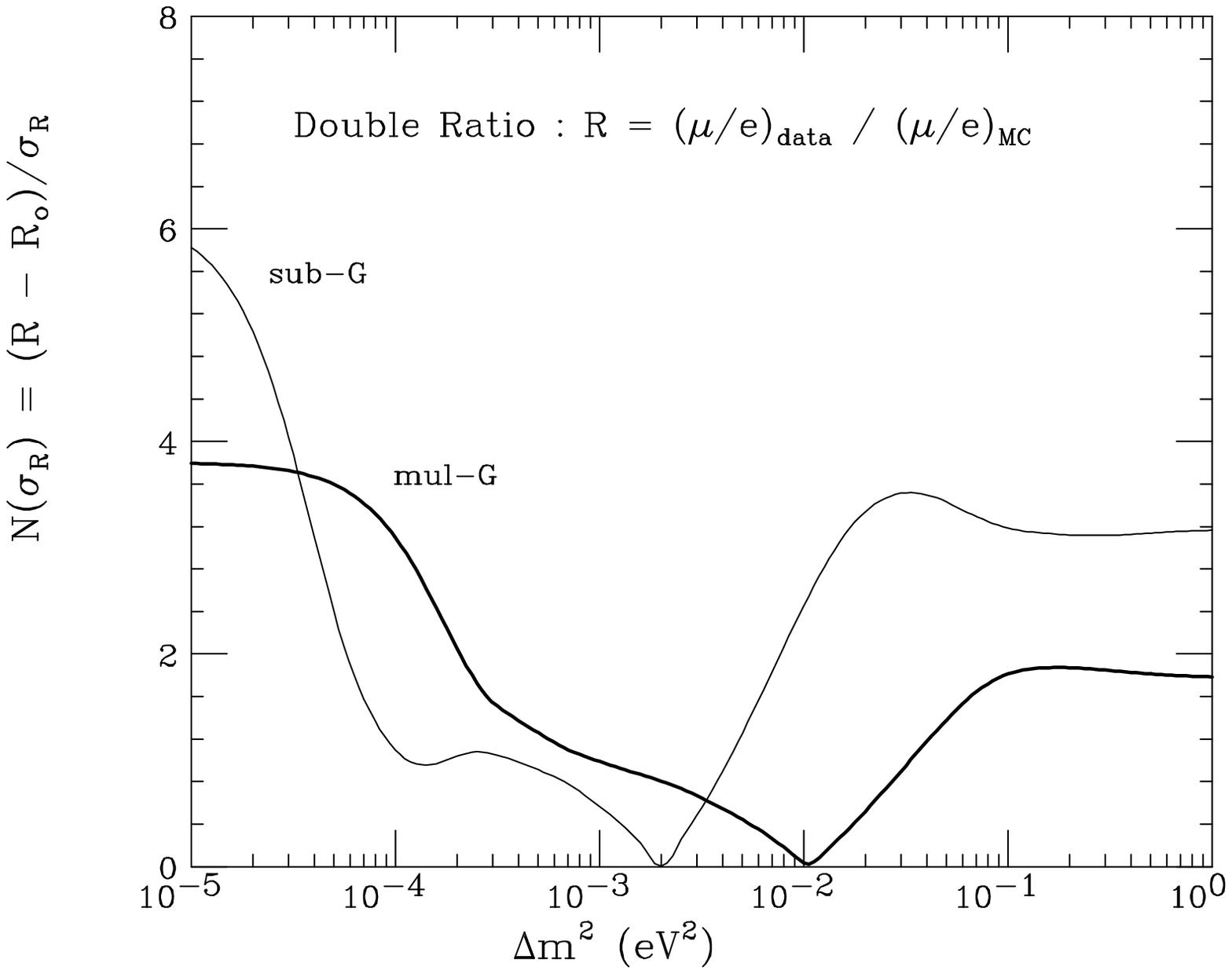,height=8.6cm}}

\vspace{0.25 cm}
\fcaption { Deviations  (in standard  deviations)
of the measured SK  sub--GeV and multi--GeV  double--ratios
 with respect to  expectations
calculated assuming the presence
of maximal  mixing $\nu_\mu \leftrightarrow  \nu_\tau$.
\label{fig:double}}
\end{figure}

In fig.~\ref{fig:double} 
we show as a function of $|\Delta m^2|$ the
deviations of the $(\mu /e)$ double ratios for the SK sub--GeV and
multi--GeV data with respect to the prediction made assuming again the
presence of $\nu_\mu \leftrightarrow \nu_\tau$ oscillations withn
maximal mixing.  To compute the number of standard deviations the
value of $\sigma_R$ has been estimated combining in quadrature the
statistical errors with the systematic uncertainty as given by SK.
The  shape of the  curves in
 fig.~\ref{fig:double} are  qualitatively easy to understand,
the suppression of the $\mu$--like events grows monotonically from
zero for very small $|\Delta m^2|$ when no neutrinos has time to
oscillate, to 0.5 at large $|\Delta m^2|$, when the effect of rapid
oscillations can be averaged.  
Since the oscillation probability is a
function of $|\Delta m^2|/E_\nu$, and the measured double ratios for
the sub--GeV and multi--GeV are approximately equal (0.67 and 0.66),
the curve for the multi--GeV data that describes the suppression of
higher energy neutrinos, is shifted to higher values of of $|\Delta
m^2|$.  Comparing fig.~\ref{fig:asym} and fig~\ref{fig:double} one can see
that the lowest value of $|\Delta m^2|$ of an allowed interval for SK
is determined by the measurement of the double ratio
for multi--GeV events.

\begin{figure} [bt]
\centerline{\psfig{figure=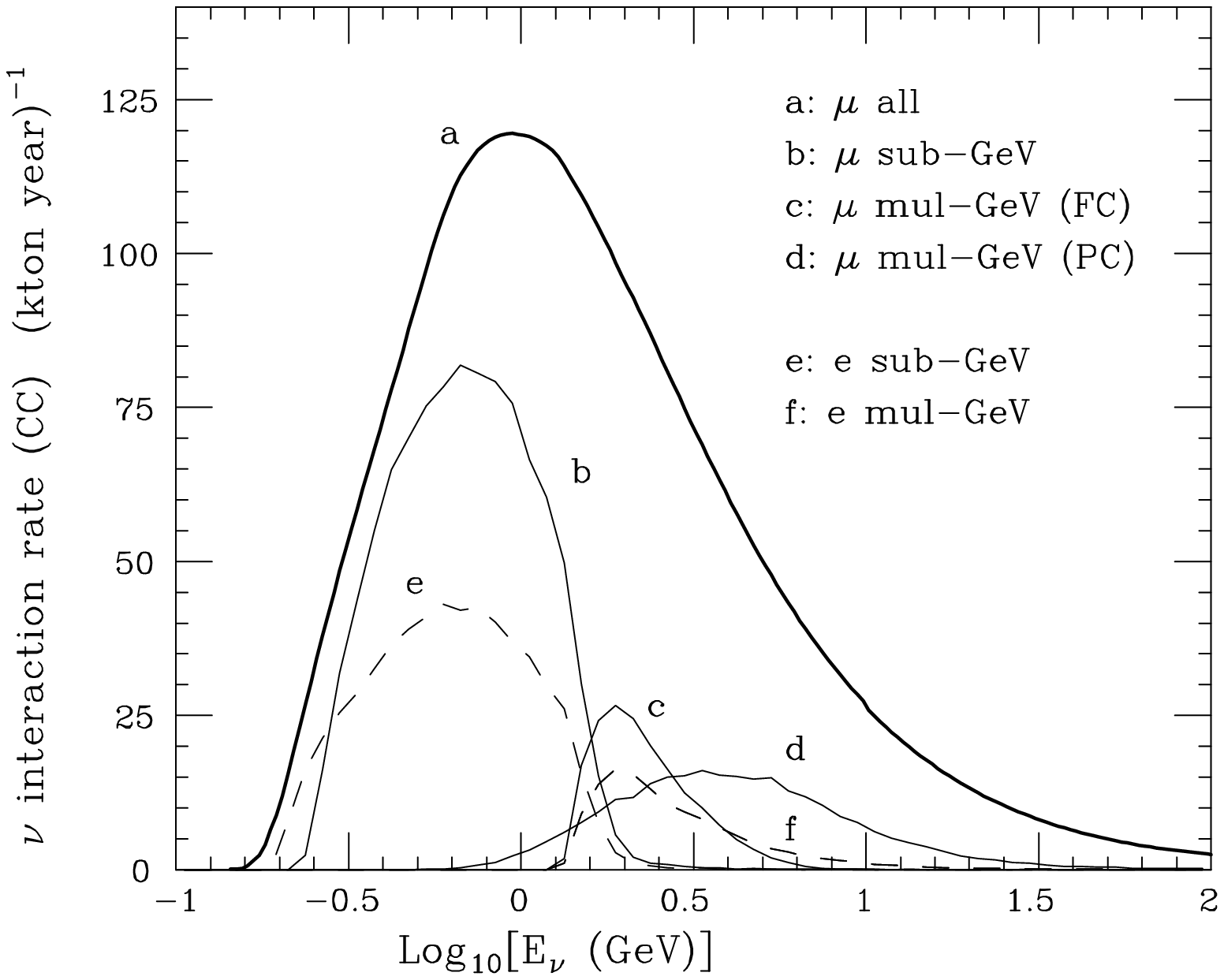,height=8.6cm}}

\vspace {0.35 cm}
\fcaption {
Approximate energy  distributions
of  interacting neutrinos  for 
different classes of events in SK. Note the difference
in  the shape of 
the response for $\mu$--like  and  $e$--like   events.
\label{fig:ener}}
\end{figure}

As in the case of Soudan, the analysis  of the double ratio
is  model  dependent, 
 because since the $e$--like and $\mu$--like events are
produced by different ranges of $E_\nu$,  different  assumptions for
the shape of the neutrino energy spectrum result in different predictions
for the ($\mu /e)$ ratio, and therefore to different interpretations
of the experimental results.
In fig.~\ref{fig:ener}
shows the approximate  energy distributions of the neutrinos that produce
different event samples in SK.  It is instructive to observe that
because of the containement requirement, in the the multi--GeV sample
fully contained (FC) $e$--like events have a broader energy
distribution (extending to higher $E_\nu$) than FC $\mu$--like events,
and in fact the MC double ratio for these events is 1.16
(significantly smaller than two).  However $\mu$--like events are also
detected as partially contained (PC) events.  For this class of events
the single ring requirement 
that is an important source of inefficiency
for  large $E_\nu$ when the average multiplicity grows,   is  removed.
For  this reason the PC
events are produced by a still broader range of $E_\nu$ that extends
beyond the energy interval for $e$--like events.  

\begin{figure} [bt]
\centerline{\psfig{figure=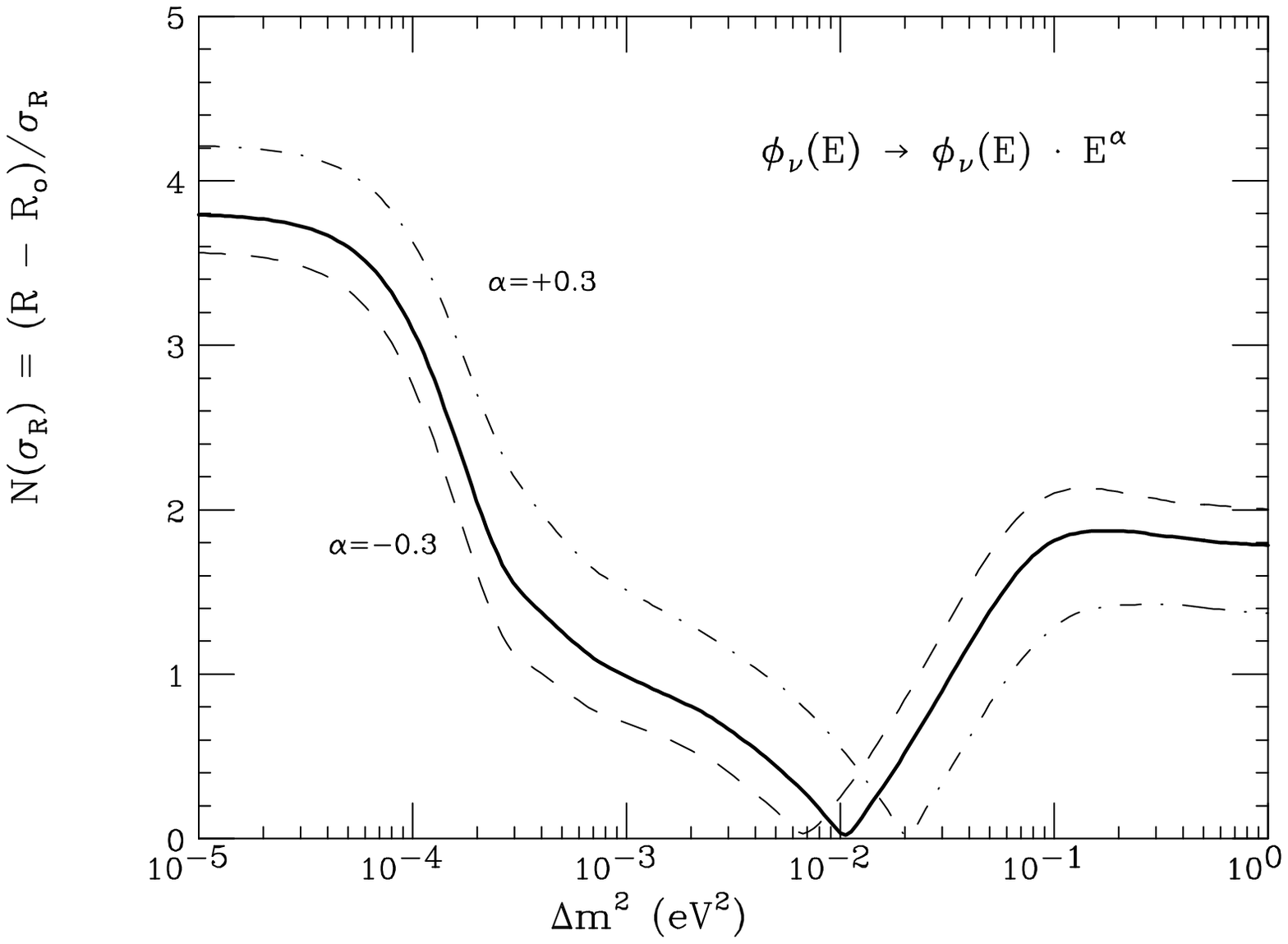,height=8.6cm}}

\vspace{0.12 cm}
\fcaption { Deviation  (in standard  deviations)
of the measured SK  multi--GeV  double--ratio with respect
to  an expectation 
calculated assuming the presence
of maximal  mixing $\nu_\mu \leftrightarrow  \nu_\tau$.
The dashed  and dot--dashed  curve are calculated
assuming  a distortion of  the HKKM $\nu$  fluxes.
\label{fig:dm_fit}}
\end{figure}

This  question is illustrated with a `pedagogical example' in
fig.~\ref{fig:dm_fit}.
  In this figure we have performed three
  Montecarlo simulations of the SK data that differ only for the
  description of the neutrino spectrum.  One calculation uses the
  standard solar minimum HKKM fluxes, while the other two assume a
  flattening (steepening) of both $\nu_e$ and $\nu_\mu$ fluxes with a
  large distortion factor $\propto E_\nu^{\pm 0.03}$ (the absolute
  normalization has no importance in this analysis).  The 
  spectral distortion has a neglible  effect on the $(\mu /e)$
  ratio at a {\em fixed} $E_\nu$, but has a significant impact on the
  predicted ratio for  the rates for the  entire multi--GeV sample
  because of  the  different    energy dependence of the
   detection efficiency for the  two  flavors.
   The flatter spectrum predicts a
  larger no--oscillation $(\mu /e)$ ratio, and in the interpretation
  of  the data in terms of $\nu_\mu \leftrightarrow \nu_\tau$
  oscillations  results  in  a  larger $|\Delta m^2|$  to explain
  the larger  suppression\footnote{An additional effect is also that 
  for  a flatter spectrum, the average energy of the neutrinos
  that  produce the  events in the sample   is  higher and  therefore
  even to produce the same  average suppression, the $|\Delta m^2|$
  must  be larger.}.

\section {Summary}
The evidence  for the disappearance of    $\nu_\mu$'s in
in  the atmospheric  neutrino   data is  robust,
and  represents  a very important result.
The detailed  interpretation of   the experimental  results
is limited  by systematic uncertainties in the prediction
of the event rates.

One of the main  sources of uncertainty   related to 
the spectrum of  primary cosmic rays,  has  been  significantly reduced
thanks to  new  measurements  with  detectors on balloons.
One more measurement   of  the primary fluxes of the 
Anti Matter  Spectrometer (AMS) should soon  become  available.
This  new  information     on the  primary fluxes  
has not yet been included  in the  simulations.

Uncertainties in the modeling of particle  production  
in hadronic interactions is  now probably the 
most important limitation  in the calculation of the 
atmospheric  $\nu$ fluxes. 
New  measurements of  the properties of particle production
in   proton--nucleus  interactions 
in the  energy range  $E_{lab} \sim 3$--100~GeV 
could  significantly  reduce  the 
systematic error   in the prediction 
 of the  normalization and  energy spectrum
of the   atmospheric  $\nu$--fluxes.
The improvement  in the determination of the primary flux
makes  this  experimental program  more  attractive  and
useful because  a  better determination of 
 the hadronic  cross sections can now  be 
directly  translated in an  improvement in the 
determination  of the atmospheric $\nu$--fluxes.

The  neutrino  cross section has also to be recognized as  an important
source of  systematic uncertainty, also relevant 
for  the LBL disappearance  programs.
The best perspectives  for  improvement
come   again from the performance of  new  experiments.

Measurements of  the muon  flux  as  a constraint 
for the  $\nu$  flux calculation are  also very important.
The results of   several measurements at high  altitude with 
detectors  on balloons  are in the process of   being analysed.
New  precise 
measurements of the muon flux at  ground level  can also
be    very important.

\section {Acknowledgements}
I would like to thank Milla  Baldo Ceolin  for the  possibility of
participating to  this  workshop, and 
Giuseppe Battistoni,
Alfredo Ferrari,
Tom Gaisser,
Maurizio Lusignoli,
Todor Stanev,
and Yoichiro Suzuki for very useful  discussions.


\begin{thebibliography}{999}
\bibitem {barbieri} This argument was  suggested to me
by Riccardo   Barbieri.
\end{thebibliography}
\end{document}